\documentclass{aa}  
\usepackage{graphicx}
\usepackage{array}
\usepackage{lscape}                                
\usepackage{natbib}
\usepackage{longtable}
\usepackage{color}
\usepackage{mathrsfs} 
\usepackage[none]{hyphenat}
\usepackage{txfonts}
\newcommand{\kms} {km\,s$^{-1}$}
\newcommand{\vsini} {$\varv$\,sin\,$i$}
\newcommand{\vmac} {$\varv_{\rm mac}$}
\newcommand{\Teff} {T$_{\rm eff}$}
\newcommand{\grav} {log\,{\em g}}
\newcommand{\fastwind} {{\sc fastwind}}
\newcommand{\msun}{M$_{\odot}$}

\begin{document} 

\sloppy

   \title{The IACOB project\thanks{Based on observations made with the Nordic Optical Telescope, operated by NOTSA, and the Mercator Telescope, operated by the Flemish Community, both at the Observatorio del Roque de los Muchachos (La Palma, Spain) of the Instituto de Astrof\'isica de Canarias.}}

   \subtitle{III. New observational clues to understand macroturbulent broadening in massive O- and B-type stars}

   \author{S.~Sim\'on-D\'{\i}az\inst{1,2},
	   M.~Godart\inst{1,2}, 
	   N.~Castro\inst{3},
	   A.~Herrero\inst{1,2}, 
	   C.~Aerts\inst{4,5}, 
	   J.~Puls\inst{6}, 
	   J.~Telting\inst{7},
	   L. Grassitelli\inst{3}
	   }

\institute{
    Instituto de Astrof\'isica de Canarias, E-38200 La Laguna, Tenerife, Spain              
    \and
    Departamento de Astrof\'isica, Universidad de La Laguna, E-38205 La Laguna, Tenerife, Spain
    \and 
    Argelander Institut f\"ur Astronomie, Auf den H\"agel 71, 53121, Bonn, Germany
    \and
    Instituut voor Sterrenkunde, KU Leuven, Celestijnenlaan 200D, 3001, Leuven, Belgium
    \and
    Department of Astrophysics/IMAPP, Radboud University Nijmegen, 6500 GL, Nijmegen, The Netherlands
    \and
    LMU Munich, Universit\"ats-Sternwarte, Scheinerstr. 1, 81679, M\"unchen, Germany
    \and
    Nordic Optical Telescope, Rambla Jos\'e Ana Fern\'andez P\'erez 7, E-38711, Bre\~na Baja, Spain.
    }

\offprints{ssimon@iac.es}

\date{Submitted/Accepted}

\titlerunning{New observational clues to understand macroturbulent broadening in massive O- and B-type stars}

\authorrunning{S. S.-D. et al.}

 
  \abstract
  {The term 'macroturbulence' of O- and B-type stars is commonly used to refer to a source of non-rotational broadening affecting their spectral line-profiles. It has been proposed to be a spectroscopic signature of the presence of stellar oscillations; however, we still lack a definitive confirmation of this hypothesis.}
  {We aim to provide new empirical clues about macroturbulent spectral line broadening in O- and B-type stars to evaluate its physical origin.}
   {We use high-resolution spectra of $\approx$430 stars with spectral types in the range O4\,--\,B9 (all luminosity classes) compiled in the framework of the IACOB project. We characterize the line-broadening of adequate diagnostic metal lines using a combined Fourier transform and goodness-of-fit technique. We perform a quantitative spectroscopic analysis of the whole sample using automatic tools coupled with a huge grid of \fastwind\ models to determine their effective temperatures and gravities. We also incorporate quantitative information about line asymmetries to our observational description of the characteristics of the line-profiles, and present a comparison of the shape and type of line-profile variability found in a small sample of O stars and B supergiants with still undefined pulsational properties and B main sequence stars with variable line-profiles (due to a well identified type of stellar oscillations or to the presence of spots in the stellar surface).}
   {We present a homogeneous and statistically significant overview of the (single snapshot) line-broadening properties of stars in the whole O and B star domain. We find empirical evidence of the existence of various types of non-rotational broadening agents acting in the realm of massive stars. Even though all of them could be quoted and quantified as a macroturbulent broadening from a practical point of view, their physical origin can be different. Contrarily to the early- to late-B dwarfs/giants, which present a mixture of cases in terms of line-profile shape and variability, the whole O-type and B supergiant domain (or, roughly speaking, stars with $M_{\rm ZAMS}$\,$\gtrsim$ 15~\msun) is fully dominated by stars with a remarkable non-rotational broadening component and very similar profiles (including type of variability). We provide some examples illustrating how this observational dataset can be used to evaluate scenarios aimed at explaining the existence of sources of non-rotational broadening in massive  stars.}
   {}


   \maketitle

\section{Introduction}\label{sec-intro}

Line profiles in optical spectra of OB stars\footnote{Along this paper, and following \cite{Ree03}, we use the term OB stars to refer to O/early-B type stars on the main sequence and their evolved descendants, the B supergiants. The remaining B-type stars are considered as a separated group, also called B dwarfs/giants.} are not only broadened by rotation. This statement has been known since the first large spectroscopic surveys of massive stars in the late 1950's \citep{Sle56,Con77}. First indirect suspicions, supported by the absence of narrow-line stars of this type, were soon confirmed with the advent of high-resolution spectrographs: the V-shape of some of the profiles did not correspond to an exclusively rotationally broadened line. More recently, the use of Fourier transform techniques \citep[cf.][]{Gra76}, in combination with profile fitting techniques, has allowed us to have access to actual projected rotational velocities (\vsini) in OB stars and quantify the relative contribution of the non-rotational and rotational broadenings \citep[e.g.,][and references therein]{Sim07, Sim14}. 

Although the extra source of line-broadening found in OB stars has been commonly quoted as macroturbulent broadening (and quantified as a macroturbulent velocity, \vmac), its connection with large scale turbulent motions of material in the line-formation region\footnote{This concept of macroturbulent broadening was initially introduced and studied in the context of cool stars \citep[see, e.g., the review by][and other historical references therein]{Gra78}. Even though the use of this term has been extended to other star domains, it does not necessarily refer to the same type of broadening mechanism or the same physical origin.} is highly improbable \citep[see, e.g.,][]{Sim10}. An alternative scenario relates this extra-broadening to the effect of stellar oscillations on the line-profiles \citep[e.g.,][]{Luc76, How04, Aer09}. In this context, not only the most commonly considered heat-driven non-radial modes, but also other spectroscopic variability phenomena identified and/or predicted in massive stars may play a role (e.g. rotational modulation, strange modes, stochastically excited non-radial modes and/or convectively driven internal gravity waves).

The presence of a variable pulsational broadening component is well known in observed line-profiles of B dwarfs/giants located in the $\beta$-Cep and SPB (slowly pulsating B-type star) instability domains \citep[e.g.][and references therein]{Aer14}. Indirect arguments presented in the last years \citep{Aer09, Sim10} indicates that this may be also the case for B Supergiants (Sgs); however, the macroturbulent-pulsational broadening connection in the whole OB star domain, a region of the Hertzsprung-Russel diagram which to-date is (by far) less explored and understood from an asteroseismic point of view, still requires direct (observational) confirmation. Indeed, even if we assume this likely connection, the exact driving mechanism (or mechanisms) of the type of oscillations which might result in the observed (macroturbulent) profiles remains undefined \citep[but see][]{Aer09, Can09, Sam10, Shi13, Sun13a, Sim15a, Aer15, Gra15a}. 

Aiming at providing a set of empirical constraints which could be used to assess the pulsational (or any other) hypothesis to explain the physical origin of macroturbulent broadening in OB stars, we started in 2008 the compilation of a high-quality spectroscopic database including multi-epoch observations of a large sample of bright (V<9) Northern O- and B-type stars. This observational material, which has now become part of the IACOB spectroscopic database \citep{Sim11a, Sim15b}, presently comprises (a) high-resolution spectra of $\approx$620 Galactic stars covering spectral types between O4 and B9 and all luminosity classes, and (b) time-series spectra with a time-span of several years and various types of time-coverage for a selected sample of targets.

In \cite{Sim10} we used part of these observations to present first empirical evidence for the existence of a correlation between the macroturbulent broadening and photospheric line-profile variations in a sample of 13~OB Sgs. In \cite{Sim14} we concentrated on the line-broadening analysis of $\approx$200 O and early-B stars to investigate the impact of other sources of non-rotational broadening on the determination of projected rotational velocities in OB stars. In this paper we benefit from a much larger and extended (in terms of spectral type coverage) spectroscopic dataset to provide new observational clues to step forward in our understanding of macroturbulent broadening in massive stars. In particular, with this work we increase and improve the available information about the single snapshot properties of this enigmatic line-broadening in the whole O and B star domain \citep[which is presently fragmentary and not necessarily homogeneous, e.g.,][]{Rya02, Duf06, Lef07, Lef10, Fra10, Bou12, Sim14, Mar14, Mar15, Mah15}. 

In this paper, we focus on the global aspects of the line-broadening properties of the large sample we composed. Subsequent work will be tuned towards time-series analysis for a sub-sample of $\approx$70\,--\,100 targets. Sections~\ref{sec-obs} and \ref{sec-tools} describe the observational dataset and the spectroscopic analysis tools we used to extract the level of line-broadening, line-asymmetry, and the spectroscopic parameters. The variety of profiles found in the sample is illustrated and discussed in Sect.~\ref{sec-resul-qualit} while the distribution of stars in the \vsini\,--\,\vmac\ and spectroscopic Hertzsprung-Russell \citep[sHR,][]{Lan14, Cas14} diagrams is discussed in Sect.~\ref{sec-resul-quantit}. 
An extensive discussion on how these observations, once combined with further information about line-profile 
skewness and variability, can help us to definitely identify the origin of the various sources of non-rotational broadening found in massive star is presented in Sect.~\ref{sec-disc}. Last, a summary of our work -- including the main conclusions -- and some future prospects are presented in Sect.~\ref{sec-summary}.

\section{Observations}\label{sec-obs}

\begin{table*} [t!]
\begin{center}
\caption{List of stars considered for this paper, including information about line-broadening and stellar parameters, and the quantity $RSk$ (relative skewness, see Eq.~\ref{eq1}). The line used to determine the line-broadening parameters is also indicated, along with its equivalent width and the signal-to-noise ratio of the adjacent continuum. Spectral classifications indicated in column 2 must be handle with care, since they come from various sources, not all of them equally reliable. EW in m\AA, \vsini\ and \vmac\ in \kms, \Teff\ in K. See Appendix~\ref{sec-append2} for a complete version of the table.}
\label{tab-t1}
\begin{tabular}{lllcccccrcccc}
\hline\hline
\noalign{\smallskip}
Target & SpC & Line & SNR$_{\rm c}$ & EW & \multicolumn{2}{c}{\vsini} & & \vmac & $RSk$ & $\sigma_{RSk}$ & log~\Teff\ & log~${\mathscr L/\mathscr L_{\odot}}$ \\
\cline{6-7}\cline{9-9}
\noalign{\smallskip}
&     &      &     &         & FT &  GOF & & GOF &  & &    & \\
\hline
        ...     &            ...   &   ...         &  ... &  ... &  ... & ... & &... & ... & ... & ... & ...\\
        HD~16582 &            B2~IV &  \ion{Si}{iii} &  183 &  138 &   9 &   9  & &        19  & 0.02 &  0.05 & 4.34 & 2.94 \\
        HD~17081 &             B7~V &  \ion{Mg}{ii} &  195 &  294 &  18 &  19  & &        24  &-0.03 &  0.02 & 4.12 & 2.40 \\
        HD~17603 &       O7.5~Ib(f) &  \ion{O}{iii} &  239 &  317 & 109 &  99  & &       115  &-0.01 &  0.02 & 4.53 & 4.21 \\
        HD~17743 &           B8~III &  \ion{Mg}{ii} &  157 &  214 &  48 &  47  & &        22  & 0.00 &  0.06 & 4.13 & 2.21 \\
        HD~18409 &          O9.7~Ib &  \ion{O}{iii} &  256 &  179 & 131 & 128  & &    $<$ 88  & 0.08 &  0.16 & 4.51 & 3.99 \\
        HD~18604 &           B6~III &  \ion{Mg}{ii} &  184 &  305 & 131 & 132  & &    $<$ 50  & 0.07 &  0.08 & 4.11 & 2.44 \\
        HD~19820 & O8.5~III(n)((f)) &  \ion{O}{iii} &  316 &  251 & 144 & 147  & &    $<$ 54  &-0.22 &  0.11 & 4.51 & 3.91 \\
        ...     &            ...   &   ...         &  ... &  ... &  ... & ... & &... & ... & ... & ... & ...\\
\noalign{\smallskip}
        \hline
\end{tabular}
\end{center}
\end{table*}

The main observational sample discussed in this paper comprises high-resolution, single snapshot spectra of 432 Galactic stars. Basically, from the 620 O- and B-type stars in the IACOB spectroscopic database\footnote{The total amount of spectra included in the IACOB spectroscopic database by November 2015 is 4560. We note the important increase in numbers compared with the last updates in the database presented in \cite{Sim15a}: 531 stars and 3705 spectra. This illustrates how productive the on-going observing runs of the project are.} as for November 2015, we discarded:  
\begin{itemize}
\item 80 stars showing clear signatures of being a double line spectroscopic binary, a multiple system or, more generally, a composite spectrum in at least one of the IACOB spectra.
\item 25 stars in which {\em all} the main diagnostic lines considered to obtain information about \vsini\ and \vmac\ (see Sect.~\ref{sec-broad-tools}) are weak, absent or present strong spectroscopic peculiarities. Among this subsample one can find stars with a spectral type earlier than O4, very fast rotators and some Oe/Be stars.
\item 83 stars having a projected rotational velocity larger than 200~\kms\ (see explanation in Sect.~\ref{sec-resul-vsvm}).
\end{itemize}
The IACOB database includes spectra from two different instruments: the FIES \citep{Tel14} and HERMES \citep{Ras11} spectrographs attached to the 2.56-m Nordic Optical Telescope and the 1.2-m Mercator telescope, respectively. Both instruments provide a complete wavelength coverage between 3800 and 7000~\AA\
(9000~\AA\ for the case of HERMES spectra), and the associated resolving power (R) of the spectra is 25000/46000 (FIES) and 85000 (HERMES). By default, all the spectra in the IACOB database are reduced using the corresponding available pipelines (FIEStool\footnote{\tt \tiny http://www.not.iac.es/instruments/fies/fiestool/FIEStool.html} and HermesDRS\footnote{\tt \tiny http://www.mercator.iac.es/instruments/hermes/hermesdrs.php}, respectively) and normalized by means of own procedures implemented in IDL. 

The best signal-to-noise ratio (SNR) spectrum per considered star was selected for the purposes of this study. The typical SNR is in the range 150-300. In many cases, especially for targets brighter than V=6, the same star was observed with both instruments. For these stars we preferred the HERMES spectrum over the FIES
one whenever the associated SNR is similar, because of the larger resolving power. 

This main (single snapshot) spectroscopic dataset is complemented with high-resolution spectroscopic time-series of a sample of 8 selected targets (Sect.~\ref{sec-resul-step-var}). Half of these stars are well known pulsators and stars with spots located in the B main sequence star domain \citep[selected from the sample described and analyzed in][and references therein]{Aer14}, the other four correspond to bright O stars and B~Sgs surveyed by the IACOB project \citep[as an extension of the observations presented in][]{Sim10}. In all cases we rely on more than 90 spectra gathered during more than 4 years. These observations are used to provide a first comparison of the type of line-profile variability present in OB stars with an important contribution of the macroturbulent broadening and B main sequence stars with well identified causes of spectroscopic variability.

\section{Tools and methods}\label{sec-tools}

In this section we describe the strategy we followed to extract from the spectra information about (a) \vsini\ and \vmac, (b) the amount of asymmetry of the diagnostic lines considered to derive the line-broadening parameters, and (c) the stellar parameters which allow us to locate the stars in the spectroscopic HR diagram. The results from the analysis, that will be used for the discussion in Sects.~\ref{sec-resul} and \ref{sec-disc}, are summarized in Table~\ref{tab-t1}, where the meaning of each column is explained below.

\subsection{Line-broadening parameters}\label{sec-broad-tools}

We applied the {\tt iacob-broad} tool \cite[][in the following SDH14]{Sim14} to the \ion{O}{iii}~$\lambda$5591, \ion{Si}{iii}~$\lambda$4552, \ion{Mg}{ii}~$\lambda$4481, or \ion{C}{ii}~$\lambda$4267 line (depending on the spectral type of the star) to derive \vsini\ and \vmac\ in the whole sample. We followed the strategy described in SDH14; namely, the line-broadening analysis is based on a combined Fourier transform (FT) + goodness-of-fit (GOF) methodology where we consider a radial-tangential definition for the macroturbulent profile, with equal radial and tangential components, and the starting intrinsic profile for the GOF computation is simplified by a $\delta$-function. 

In most cases, the outcome of the global rectification applied to all IACOB spectra was fairly good when zooming into the smaller spectral windows associated with the lines to be analyzed. However, whenever necessary, we activated the option available in {\tt iacob-broad} to locally improve the normalization before performing the line-broadening analysis.

\begin{figure}[!t]
\centering
\includegraphics[height=0.47\textwidth, angle=90]{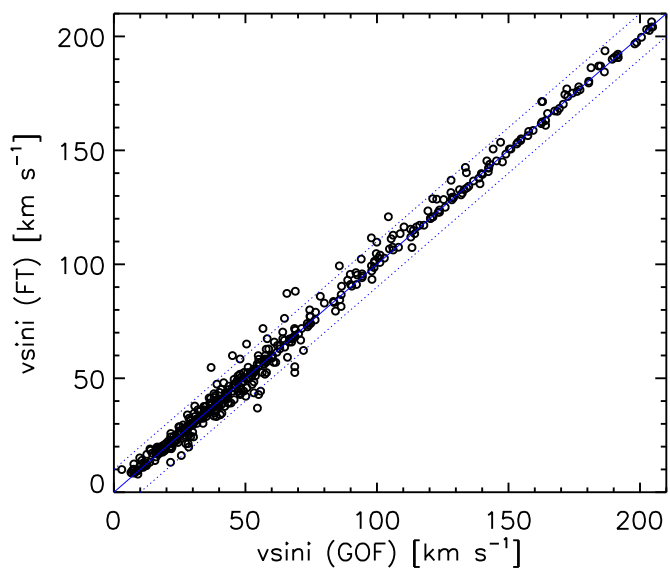}
\caption{Comparison of the \vsini\ values determined by means of the FT and GOF methodologies using the {\tt iacob-broad} tool. The agreement is better than $\pm$~10~\kms\ for 95\% of the sample, and better than $\pm$~20~\kms\ for the remaining stars.}
\label{fig-f1}
\end{figure}

Results from the line-broadening analysis of the 432 stars are indicated in columns 6\,--\,8 of Table~\ref{tab-t1} along with some information about the equivalent width (EW, column 5) of the considered diagnostic line (column 3) and the SNR of the adjacent continuum (column 4). We provide both FT and GOF solutions for \vsini, but only the latter, together with the associated \vmac\ will be considered in this paper. The main reason is that we found a general good agreement between the \vsini\ derived by means of both methods (better than $\pm$10 \kms\ in most cases, see Fig.~\ref{fig-f1}), but (a) the GOF approach is less affected by the subjectivity in the selection of the first zero of the Fourier transform in some complicated cases, and (b) the GOF solution allows to detect cases where we can only provide upper limits to \vmac. The latter refers to situations in which the \vmac\ value associated with the best fitting solution (minimum $\chi^2$) is larger than zero, but \vmac\,=\,0 is still an acceptable solution (below the 1-$\sigma$ confidence level). As we show and discuss in Sect.~\ref{sec-resul}, these cases normally correspond to profiles in which the rotational broadening component dominates. Those cases in which \vmac\ could not be properly determined are quoted in Table~\ref{tab-t1} by providing the value corresponding to the best fitting solution as an upper limit.

\subsection{Line-profile asymmetry}\label{sec-asym-tools}

Most of the previous work on macroturbulent broadening in OB stars concentrate on the characterization of the line-profiles via \vsini\ and \vmac. However, as is well known in the context of stellar oscillations \citep[cf.\ Chapters 5 and 6 of][]{Aer10b}, such a simplistic two-parameter description lacks an important piece of information about the properties of the profiles: the amount of {\it line-asymmetry\/}. A particularly illustrative study in this respect is the one by \cite{Sch97}, where the predicted variability of line-profiles
due to adiabatic non-radial pulsations in rotating stars is shown for a large variety of oscillation and rotation parameters. This reference work contains various diagnostics used to describe the snapshot characteristics of the line-profiles, a combination of which we adopt here as the {\it relative skewness} ($RSk$). We define this quantity as
\begin{equation}\label{eq1}
RSk\,\equiv\,\langle \varv^3 \rangle  / \langle \varv^2 \rangle^{3/2}, 
\end{equation}
where
\begin{equation}\label{eq2}
\langle \varv^n \rangle\,=\,\frac{\int^{\infty}_{-\infty} (\varv - \langle \varv \rangle )^n  (1-F(\varv)) d\varv}{\int^{\infty}_{-\infty} (1-F(\varv)) d\varv}  \hspace{2cm} {\rm for} \hspace{0.2cm} n=2, 3
\end{equation}
are the second and third normalized central moments of a spectral line denoted as ($\varv$, $F(\varv)$), adopting the definition used in \cite{Sch97}, and
\begin{equation}\label{eq3}
\langle \varv \rangle\,=\,\frac{\int^{\infty}_{-\infty} \varv~(1-F(\varv)) d\varv}{\int^{\infty}_{-\infty} (1-F(\varv)) d\varv}.
\end{equation}
The dimensionless quantity\footnote{In statistical terms, this quantity is quoted as third standardized moment or (Pearson's) moment coefficient of skewness.} defined by Equation~(\ref{eq1}) represents the amount of skewness of the line-profile relative to its total width. It allows to compare the amount of asymmetry for lines with very different widths in a meaningful way. 

We used the same line-profiles considered to derive the line-broadening parameters (Sect.~\ref{sec-broad-tools}) for the computation of the first three moments and, ultimately, the quantity $RSk$. To compute the latter, we considered various integration limits and found the best overall choice to be an integration centered on $\langle \varv \rangle$ and extending over $\pm$2.5$\sqrt{\langle \varv^2 \rangle}$. These integration limits were selected as the best compromise between including as much information as possible from the extended wings of the V-shaped line-profiles while avoiding spurious subtleties associated with the noise of the adjacent continuum.

The computed values of $RSk$ for the whole sample of stars, along with their corresponding uncertainties, are presented in Columns 9 and 10 in Table~\ref{tab-t1}. These uncertainties result from the formal propagation of errors for Eqs~(\ref{eq1})--(\ref{eq3}), assuming that the only source of uncertainty is the noise associated with the normalized flux of the line-profile, which is inversely proportional to the signal-to-noise ratio of the adjacent continuum (SNR$_{\rm c}$).  Given the stability of the spectrographs and the level of precision of the wavelength calibration, this approximation is fully justified.

\subsection{Stellar parameters}\label{sec-param-tools}

The spectroscopic stellar parameters of the sample were obtained by means of {\tt iacob-gbat} \citep[][O stars]{Sim11b} or an updated version of the tool described in \citet[][B stars]{Cas12}. These tools, aimed at performing quantitative spectroscopic analyses of O and B-type stars in a fast, automatic way, are based on a huge, pre-computed grid of \fastwind\ \citep{San97, Pul05, Riv12} synthetic spectra and a $\chi^2$ minimization line-profile fitting technique. Some notes on the followed strategy can be found in \cite{Lef07}, \cite{Cas12}, and \cite{Sab14}. In this case, we fixed \vsini\ and \vmac\ to the values resulting from the line-broadening analysis described above. From the whole set of output parameters provided by the automatic tools, we concentrate on the derived \Teff\ and \grav\ because they allow to locate the studied stars in the sHR diagram. 

We provide log~\Teff\ and log~${\mathscr L/\mathscr L_{\odot}}$\footnote{${\mathscr L}$\,:=\,\Teff$^{4}$ / g $\sim$ $L$/$M$ $\sim$ $\Gamma_{\rm Edd}$ $\sim$ $g_{\rm rad}/g_{\rm grav}$ \citep{Lan14}.} in the two last columns of Table~\ref{tab-t1} to facilitate the identification of stars in the corresponding diagrams. In addition, we highlight below some important points regarding the compiled set of stellar parameters:
\begin{itemize}
\item Given the quality of the spectroscopic observations and the strategy we have followed for the spectroscopic analysis, we can assume $\approx$5\% and $\approx$0.15~dex as rough estimations for the uncertainties in the derived \Teff\ and \grav, respectively.
\item There are $\approx$50 stars for which we do not provide stellar parameters in Table~\ref{tab-t1}. Most of them are late-B stars whose parameters lay outside our grid of \fastwind\ models. The lower \Teff\ boundary of the grid is 11000 K and this implies that our analysis tools do not provide reliable parameters for stars with \Teff$\le$12000~K.
\item We have not checked one-by-one all the analyzed spectra in detail. Therefore, the stellar parameters quoted in Table~\ref{tab-t1} must be considered with caution for purposes other than those discussed in this paper, especially concerning the investigation of individual stars.
\end{itemize}

\section{Results}\label{sec-resul}

\subsection{Line profiles in O and B stars: a qualitative overview}\label{sec-resul-qualit}

\begin{figure}[!t]
\centering
\includegraphics[width=0.47\textwidth]{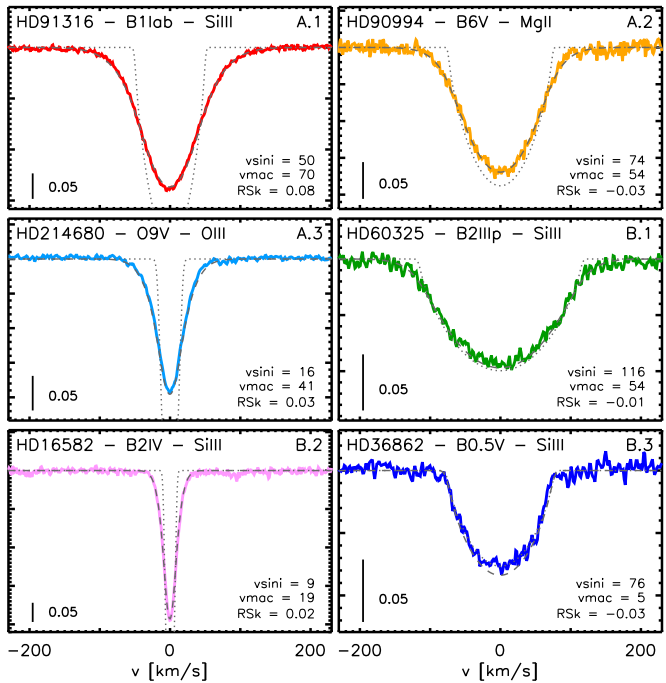}
\caption{Illustrative examples of the various types of line-profiles found in the IACOB sample of O- and B-type stars. The values of \vsini\ and \vmac\ resulting from the line-broadening and line asymmetry analysis (see
Sect.~\ref{sec-tools}) are quoted in the lower right corners. Each profile has been selected to be a representative of the six regions indicated in Fig.~\ref{fig-f4} (the same color and A-B label codes are used here). The profiles are organized from bottom to top, and from left to right, following an increasing sequence of \vmac\ and \vsini, respectively.
}
\label{fig-f2}
\end{figure}

\begin{figure}[!t]
\centering
\includegraphics[width=0.47\textwidth]{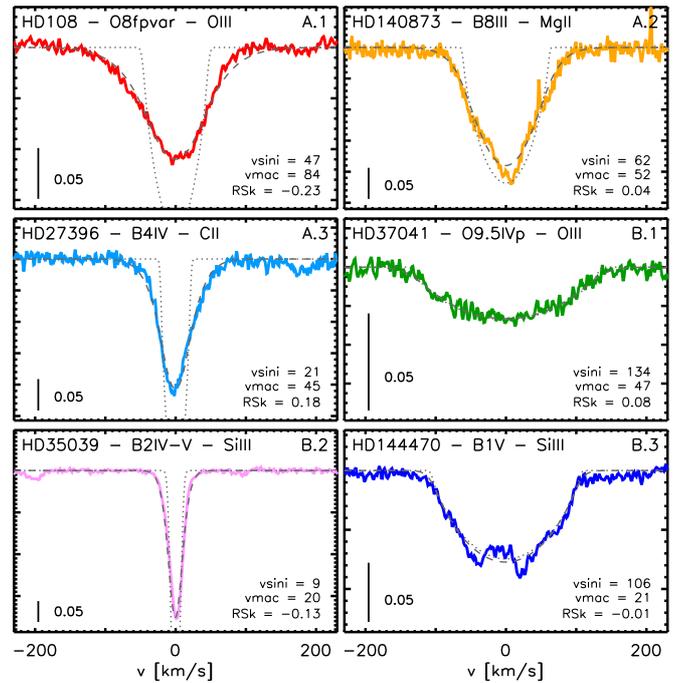}
\caption{Same as Fig.~\ref{fig-f2} but for line-profiles which are more asymmetric (A.1, A.3, B.2) or have a more complex structure (A.2, B.3). A case for which only an upper limit of \vmac\ can be obtained is also illustrated in B.1. In the latter, {\tt iacob-broad} gives (\vsini, \vmac)\,=\,(134, 47) -- dashed line -- as the best fitting solution, but (134, 0) -- dotted line -- is an equally acceptable solution.}
\label{fig-f3}
\end{figure}

Figures~\ref{fig-f2} and \ref{fig-f3} show some representative examples of the various types of line-profiles found in stars in the IACOB sample. In all cases the observed profiles, colored and labeled following the guidelines described in Sect.~\ref{sec-resul-vsvm}, are complemented with information from the outcome of the line-broadening and line-asymmetry analysis (see Sect.~\ref{sec-tools}). We also over-plot, for guidance purposes, synthetic profiles with the same equivalent width as the observed ones, convolved with the indicated \vsini\ (dotted lines) and \vsini\,+\,\vmac\ values (dashed lines). The comparison between the dotted lines and the observed profiles allows us to visualize the effect of the non-rotational broadening on the shape of the lines for different \vsini\,-\,\vmac\ combinations. The dashed lines can be used to assess the quality of the final fit and visually identify line-profiles which are clearly asymmetric or have a bumpy shape (see e.g., Figure~\ref{fig-f3} and description below). 

As illustrated by Figures~\ref{fig-f2} and \ref{fig-f3}, the width of the profiles ranges from narrow to broad, and its shape from clearly roundish, as expected in a rotationally dominated case, to triangular, where the so-called macroturbulent broadening is dominating. Many different combinations of width and shape occur. Most of the profiles are smooth (i.e., do not have any detectable substructure) and symmetric from visual inspection, having similar characteristics as those shown in
Figure~\ref{fig-f2}; however, there is also a non-negligible number of stars showing asymmetric profiles and/or spectroscopic signatures which could be associated with the effect of a certain type of pulsations on the line
profiles\footnote{The most clearly detectable cases from single snapshot spectra are those associated with coherent pressure modes as in $\beta$~Cep stars with moderate projected rotational velocities \citep[see, e.g., Panel B.3 in Figure~\ref{fig-f3} and][for more illustrative examples]{Tel06} or coherent gravity modes that occur in so-called SPBs \citep{Dec02}.}, spots, a magnetosphere, and/or undetected binarity. Some of these cases are shown in Figure~\ref{fig-f3}.

\subsection{Line-broadening in O and B stars: a quantitative overview}\label{sec-resul-quantit}

Outside the asteroseismology community, it is common to summarize the information about the width and shape of spectral line-profiles by means of only two time-independent line-broadening parameters: \vsini\ and \vmac. While the first one has a well defined physical meaning, \vmac\ is only a tuning parameter used to quantify the effect of any other broadening mechanism which is not rotation. For example, as indicated by \cite{Aer14}, periodic line-profile variability caused by surface inhomogeneities or by oscillations in main-sequence B stars can be mimicked by a combination of time-dependent rotational and macroturbulent broadening. Moreover, the value derived for this parameter depends on specific assumptions on the definition of the macroturbulent profile (e.g. radial-tangential vs. isotropic Gaussian, percentage of radial vs. tangential components). In addition, as highlighted in SDH14 \citep[see also][]{Sun13b, Mar14}, there is empirical evidence indicating that the methodology described in Sect.~\ref{sec-broad-tools} for disentangling rotation from other sources of line-broadening may be failing in some specific cases. This mainly refers to potential limitations -- of still unclear origin -- in having access to the actual value of the projected rotational velocity in stars with line-profiles dominated by the so-called macroturbulent broadening.
Therefore, we must handle any quantitative interpretation of the macroturbulent 
broadening in O and B stars in terms of \vmac\ with care, especially when combining measurements from different sources which might not be using the same techniques or assumptions.

A detailed investigation of the shortcomings of the application of the FT and GOF methodologies in the whole O and B star domain -- following some guidelines presented in SDH14 -- is planed for a subsequent paper of this series (see also some notes in Sect.~\ref{sec-resul-step-var}). In the meanwhile, we rely on the measurements resulting from the application of the methods described in Sect.~\ref{sec-broad-tools} to provide a complete homogeneous overview of the single snapshot line-broadening characteristics of stars in the whole O and B star domain. 
Despite some of the \vsini\ and \vmac\ values quoted in Table~\ref{tab-t1} may change in the future, the main conclusions presented in these sections will remain since the followed approach provide a representation of the global shape of the line-profiles valid enough for the purposes of this work. 

\begin{figure}[!t]
\centering
\includegraphics[height=0.47\textwidth, angle=90]{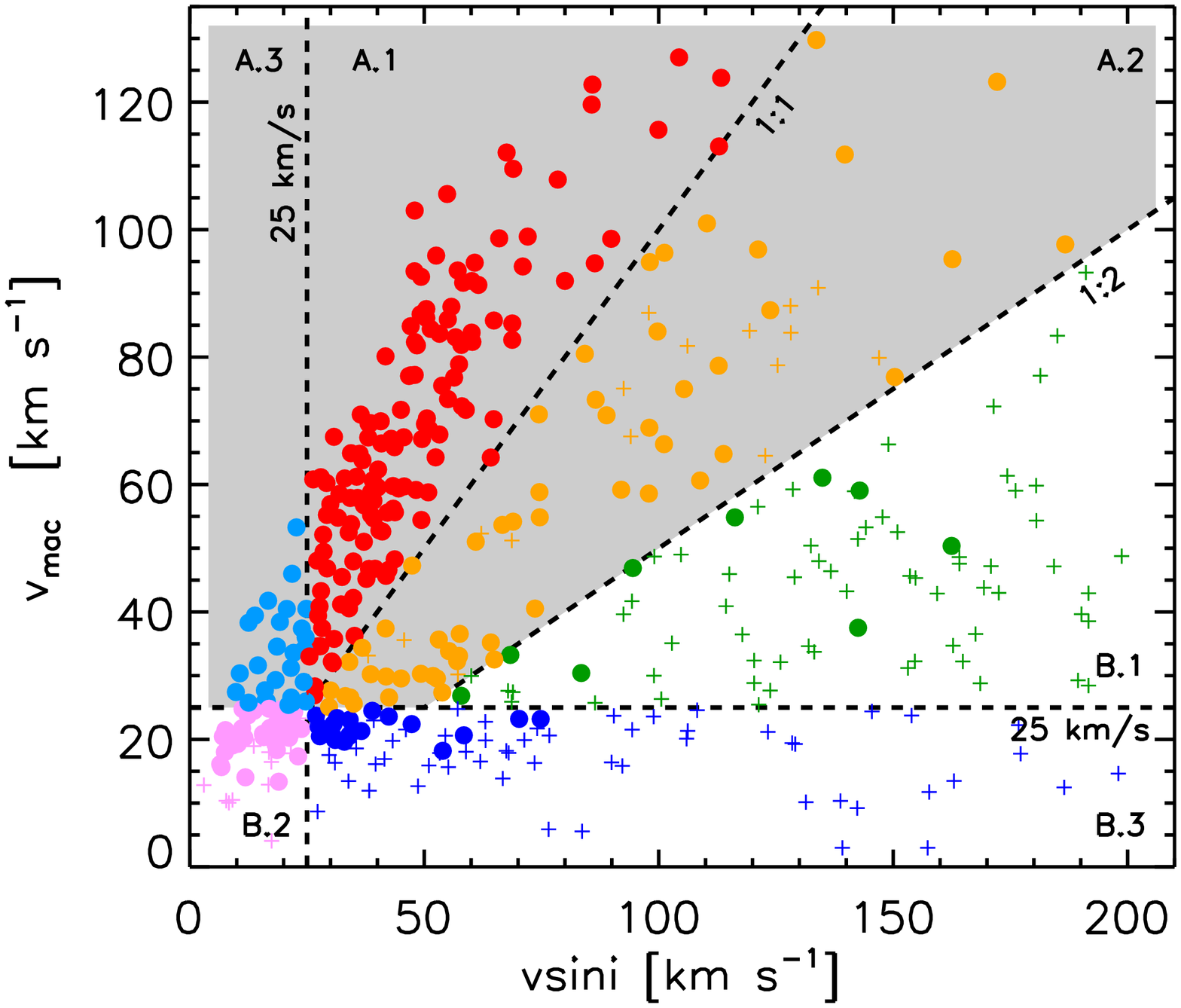}
\caption{Line-broadening characterization via \vsini\ and \vmac\ of a sample of 432 Galactic O and B stars. Crosses correspond to stars for which only an upper limit in \vmac\ could be determined. Various colors and lines separate stars affected by a different relative rotational and macroturbulent broadening contribution.  The 25~\kms\ horizontal and vertical lines indicate conservative limitations of the performed methodology (see main text for explanation). Gray-shadowed region groups stars with an important macroturbulent (compared to rotational) broadening contribution. 
}
\label{fig-f4}
\end{figure}

\subsubsection{Distribution of stars in the \vsini\,--\,\vmac\ diagram}\label{sec-resul-vsvm}

The results from the line-broadening analysis of our sample of stars in the \vsini\,-\,\vmac\ diagram are presented in Fig.~\ref{fig-f4}. As previously found by \cite{Lef10}, \cite{Mar14} and SDH14 using more specific samples (in terms of spectral type and luminosity class), the whole diagram is populated below \vmac$\approx$120~\kms, except for a gap found in the low \vsini\ regime, where stars with \vmac/\vsini ~$\gtrsim$~2 and \vmac~$\gtrsim$~40~\kms\ are absent. This absence may be due to the limitations of the considered methodology (see notes in SDH14).  The diagram has been divided into six regions depending on the
relative contribution of the rotational and macroturbulent broadening and taking into account a conservative lower limit of reliability in the \vsini\ and \vmac\ measurements of 25~\kms\ (see SDH14 for more details). Figures~\ref{fig-f2} and \ref{fig-f3} show some illustrative examples of the type of profiles found in each of these regions.

We separate the sample in two broad groups: stars showing an important (or dominant) macroturbulent broadening contribution are marked in red (A.1), orange (A.2), and cyan (A.3) and grouped by the gray shadow region; the other three regions consist of stars in which either rotational broadening dominates (green and dark blue, B.1 and B.3, respectively) or both \vsini\ and \vmac\ are below the limits of reliability of the methodology (pink, B.2). While there are many stars in this latter box for which \vmac~$>$~\vsini, the non-rotational
broadening could actually be associated with so-called microturbulence (SDH14) or with known sources of pulsational broadening \citep[see][]{Aer14}. We hence exclude these stars from those with clearly dominated macroturbulent profiles.

\begin{figure*}[!t]
\centering
\includegraphics[width=0.42\textwidth, angle=90]{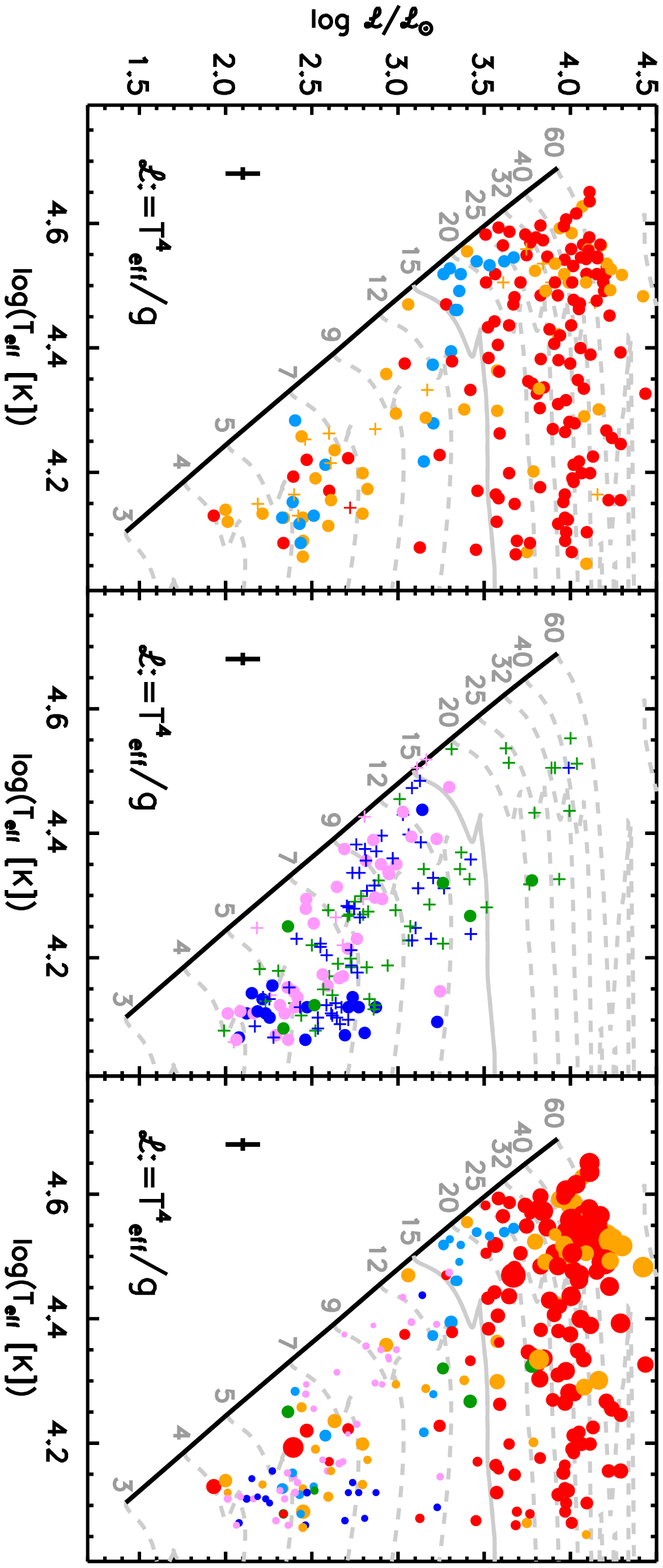}
\caption{Location of the stars from groups A [left panel] and B [middle panel], as defined in Sect.~\ref{sec-resul-vsvm}, in the sHR diagram. Same color code as in Fig.~\ref{fig-f4}. The Geneva, non-rotating evolutionary tracks \citep{Eks12} are over~plotted for guidance purposes. The black cross to the left shows the typical uncertainties in log~\Teff\ and log~${\mathscr L}$  (assuming conservative uncertainties in \Teff\ and \grav\ of 5\% and 0.15~dex, respectively, and no correlation between both quantities). [Right panel] All stars from groups A and B (except those for which \vmac\ could not be properly determined) joined together with symbol sizes proportional to \vmac.
}
\label{fig-f5}
\end{figure*}

We indicate with crosses those stars for which the {\tt iacob-broad} analysis was only able to provide upper limits for the value of \vmac\ (see notes in Sect.~\ref{sec-broad-tools}). Panel B.1 in Figure~\ref{fig-f3} shows one of these cases. Most of the stars are concentrated in the region where \vmac/\vsini~$\le$~0.5.  This is expected since the extra-broadening only produces a small effect on the wings of the line-profile when rotational broadening dominates. Although they are not shown in Fig.~\ref{fig-f4}, we found that this situation extends to all stars in the IACOB sample with \vsini$>$200~\kms. Therefore, we decided to exclude them from the discussion below after checking that this decision does not modify any of the conclusions presented in this paper\footnote{For information, the 83 excluded stars with \vsini$>$200 \kms\ have a very similar distribution in the spectroscopic HR diagram as that shown in the middle panel of Fig.~\ref{fig-f5} by the green and dark blue crosses. The same comment applies to Fig.~\ref{fig-f7}.}.

Despite the multi-epoch character of the IACOB spectroscopic dataset, there may still be some cases of undetected binarity or composite spectra.  We have only one spectrum for some of the stars in the database and some composite spectra may not be detected with the (minimum 3) spectra we have for most of the targets. Some stars are already marked as suspicious in view of the shape of the line-profiles (e.g., HD~140873 in Fig.~\ref{fig-f3}); however, some others can still remain unidentified,especially when we only have one spectrum. We hence must keep in mind that there may be some cases in which the single snapshot line-broadening analysis presented here could end up in spurious results. In particular, a composite spectrum with two lines that are not clearly separated can be artificially fitted with a line-profile having a larger \vsini\ and, specially, larger \vmac\ than any of its individual components.

\subsubsection{Macroturbulent broadening across the upper part of the sHR diagram}\label{sec-resul-sHR}

The location of the stars from groups A and B in the sHR diagram is presented separately in left and middle panels, respectively, of Fig.~\ref{fig-f5}. We use the same colors and symbols as in previous figures. The non-rotating evolutionary tracks for solar metallicity from \citet{Eks12} are also shown for guidance purposes. The typical uncertainties in log~\Teff\ and log~${\mathscr L}$ (see Sect.~\ref{sec-param-tools}) are indicated with a black cross in the bottom left corner of each panel.  These two panels are complemented with a third one which shows the same distribution of stars but from a different perspective. This time, we plot all stars together (except those for which \vmac\ could not be properly determined, i.e. the cross symbols in Figs.~\ref{fig-f4} and \ref{fig-f5}), using the same color code, but scaling the size of the symbols with \vmac. 
All together, these figures present, for the first time, a homogeneous and statistically significant overview of the (single snapshot) line-broadening properties of stars in the whole O and B star domain.

We highlight below the main results that can be extracted from inspection of Fig.~\ref{fig-f5}, regarding the distribution of stars in the sHR diagram in terms of line-profile shape characteristics:
\begin{itemize}
\item The upper part of the sHR diagram -- including the O stars and B~Sgs or, roughly speaking, stars with initial masses $\gtrsim$15~M$_{\odot}$ -- is basically populated by stars with a remarkable non-rotational broadening component (red, orange and cyan circles). In this region of the diagram, there are only a few stars from group B (mostly green crosses with \vsini\,$>$130~\kms).
\item The relative number of stars with profiles dominated by macroturbulent broadening (red circles) becomes much smaller when moving down in the sHR diagram. There are a few outliers in the lower part of the diagram having large values of \vmac; however, these may actually be non-detected spectroscopic binaries (see note in Sect.~\ref{sec-resul-vsvm}). Indeed, new observations obtained during the development of this paper have allowed us to confirm this hypothesis in a couple of targets (now excluded from the final sample). Other similar cases may still remain in the sample.
\item Contrarily to the homogeneity -- in terms of line-broadening properties -- found in O stars and B~Sgs, the situation is much more heterogeneous in the B dwarf/giant region ($\lesssim$~15~M$_{\odot}$). The only common characteristic in this lower part of the sHR diagram is that the amount of non-rotational broadening (quantified as \vmac) is much smaller in absolute terms in comparison with stars with larger masses (see right panel in Fig.~\ref{fig-f5}). Apart from this, below log~$\mathscr{L}/\mathscr{L}_{\odot}~\sim$~3.5, stars with different types of profiles are not concentrated in specific/separated regions of the diagram.  
\end{itemize}
We interpret these results as empirical evidence of the existence of various types of non-rotational broadening agents acting in the realm of massive stars. Even though all of them can be quantitatively characterized in terms of \vmac\ -- and quoted as macroturbulent broadening -- from a practical point of view, their physical origin can be different. Indeed, it is natural to think that some of them could be acting at the same time, with a different relative contribution to the line-profiles, depending on the specific properties of the star at a given moment during its evolution. 

Under this scenario, the distribution of stars shown in Fig.~\ref{fig-f5} indicates that, below $\approx$15~M$_{\odot}$ (or, equivalently log~$\mathscr{L}/\mathscr{L}_{\odot}~\sim$~3.5) the final shape of the stellar lines depends on a combination of different factors which are not only controlled by \Teff, \grav, and/or \vsini. This statement is in agreement with \citet[][see also notes in Sect.~\ref{sec-resul-step-var}]{Aer14}, who showed that, in the B main sequence star domain, the presence of stellar spots, surface inhomogeneities, and stellar oscillations are effective sources of (variable) line-broadening. Therefore, in this case, we should avoid using the term macroturbulent broadening, or at least, be aware that the sources of broadening in these stars may have nothing to do with what we call macroturbulent broadening in other regions of the HR diagram (e.g. the Sun, cool stars, A-type stars, or even O stars and B~Sgs). As indicated by \cite{Aer14} the correct identification of the broadening agents should not rely on single snapshot, but on multi-epoch observations.

\begin{figure}[!t]
\centering
\includegraphics[width=0.48\textwidth]{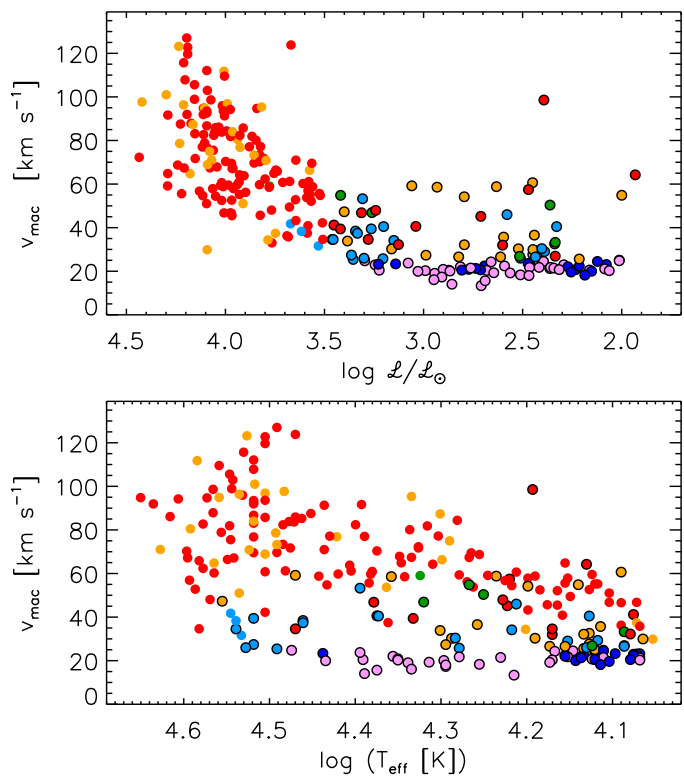}
\caption{Dependence of \vmac\ with log~${\mathscr L}$ and \Teff\ for the same stars included in the right panel of Fig.~\ref{fig-f5}. Stars with log~$\mathscr{L}/\mathscr{L}_{\odot}$\,$<$\,3.5 are marked in both panels with open black circles for better identification.}
\label{fig-f5b}
\end{figure}

Above $\approx$15~M$_{\odot}$ we find a more homogeneous distribution of profile types, suggesting one main broadening agent in the upper part of the sHR diagram, whose relative contribution becomes gradually less important for lower masses. Since very similar profiles are found for stars covering a broad range of effective temperatures, gravities, rotational velocities, and wind properties, none of these stellar properties seems to be a key factor in the occurrence of this extra-broadening. In addition, the fact that this broadening agent is operating in a very similar way for stars in different evolutionary stages (from the O dwarf to the late-B supergiant phases) can be considered as a strong empirical constraint to any scenario trying to explain its physical origin.

Figure~\ref{fig-f5b} reinforces these statements and also add -- in a more quantitative way -- interesting information about the dependence of \vmac\ with log~${\mathscr L}$ and \Teff. Again, both panels of Fig.~\ref{fig-f5b} indicate the absence of a remarkable connection between \vmac\ and any of these two quantities for stars with log~$\mathscr{L}/\mathscr{L}_{\odot}\lesssim$3.5 (highlighted using black open circles). Above this value, where we postulate that only one main broadening agent (in addition to rotation) begins to dominate over the rest, a clear positive correlation between \vmac\ and log~$\mathscr{L}$ is detected. Since, the quantity $\mathscr{L}$ is proportional to $L$/$M$ or, equivalently, to $\Gamma_{\rm Edd}$ or $g_{\rm rad}/g_{\rm grav}$ \citep[see][]{Lan14}, one could argue that \vmac\ becomes larger when the impact of radiation pressure (at least in the outer, but still subsonic envelope) increases. However, as we discuss in Sect.~\ref{sec-resul-obs-broad}, other effects may also explain this  correlation \citep[see also][]{Gra15a}.

Circles not surrounded in black in the bottom panel of Fig~\ref{fig-f5b} show the dependence of \vmac\ with \Teff\ for the case of O stars and B~Sgs (cf., stars having log~$\mathscr{L}/\mathscr{L}_{\odot}$\,$>$\,3.5 in Fig.~\ref{fig-f5}). In addition to the decrease of \vmac\ from the early- to the late-B supergiants already pointed out by previous studies \citep[see, e.g.,][]{Duf06, Lef07, Mar08, Fra10, Mar14, Sim14}, the figure seems to indicate that there is also a small decrease of this quantity from the early-B/late-O~Sgs (log\Teff~$\sim$~4.5) towards the hotter O giants and dwarfs (log\Teff~$\gtrsim$~4.5). Last, it is interesting to note that late-O dwarfs are characterized by having the lowest values of \vmac\ among the luminous OB stars (see cyan circles with log\Teff$\gtrsim$~4.45 and log$\mathscr{L}/\mathscr{L}_{\odot}$~$>$~3.2 in Figs.~\ref{fig-f5} and \ref{fig-f5b}).

\section{Discussion}\label{sec-disc}

\subsection{On the pulsational origin of macroturbulent broadening in OB stars}\label{sec-resul-hypot}

\cite{How04} presented a summary of different scenarios which had been proposed by that time to explain the existence of what we traditionally call macroturbulent broadening in O stars and B~Sgs\footnote{In this section we mainly concentrate in the OB star domain, i.e., stars with M$_{\rm ZAMS}\gtrsim$~15~\msun.}. In brief, these scenarios invoke the effect of stellar winds, pulsations, differential rotation, turbulent motions and rotationally-induced tangential turbulence on the stellar lines. In all cases, there is an implicit assumption that the occurrence of this spectroscopic feature is associated with large-scale velocity fields globally affecting the line-formation region. 

From all these scenarios, macroturbulent broadening in the OB star domain being a spectroscopic signature of stellar oscillations is presently the most strongly considered hypothesis. Although this possibility has not been definitely proven from an observational point of view, the works by \cite{Aer09, Aer14, Sim10, Sim15a, Aer15}, and \cite{Gra15a} have produced encouraging results in this direction. 

\cite{Aer09} revived the pulsational scenario, proposed already a long time ago by \cite{Luc76}, showing that the shape of the metal line-profiles observed in the high-resolution spectra of B~Sgs can be naturally reproduced by the combined velocity broadening effect of hundreds of low-amplitude non~radial gravity-mode pulsations. This work served to support theoretically, for the first time, stellar oscillations as a viable explanation for macroturbulent broadening in OB stars.

\cite{Sim10} provided for the first time firm observational evidence for a strong correlation between the amount of non-rotational broadening affecting the line-profiles in a sample of 13 OB~Sgs and the peak-to-peak amplitude of variation of the third moment (skewness) of the lines. Interestingly, very similar trends had been obtained one year before from the simulations by \cite{Aer09}. This encouraging result helped to further support -- this time from the observational point of view -- the hypothesis of stellar oscillations being the most probable physical origin of macroturbulent broadening in B~Sgs.

In this context, we remark that \citeauthor{Aer09} did not consider the excitation of the modes but rather studied the impact of a certain type of oscillations on the line-broadening and line-profile variations.
In particular, \cite{Aer09} highlighted the importance of considering a large range of mode degrees (i.e. leading to a dense frequency spectrum of modes as, for example, occurs in gravity-mode pulsators) to reproduce, thanks to their collective effect, the observed characteristics of the macroturbulent profiles in terms of global shape and structure.

Heat-driven non-radial coherent pressure (p-) or gravity (g-) modes excited by the $\kappa$-mechanism in the iron opacity bump are clear candidates to produce stellar oscillations with these characteristics in certain regions of the HR diagram. However, these are not the only possibility. Stochastically-excited oscillations giving rise to a whole spectrum of waves driven by turbulent pressure fluctuations initiated in inefficient (in transporting energy) sub-surface convection zones \citep{Gra15a, Gra15b} or by the interface of the convective core and the radiative envelope \citep{Rog13, Aer15} are other options that have been proposed. All these causes (plus maybe also others not mentioned here, e.g. strange modes, especially in the high $L/M$ regime) could be acting together, contributing with a different weight depending on the specific star.

\begin{figure*}[!t]
\centering
\includegraphics[width=0.47\textwidth]{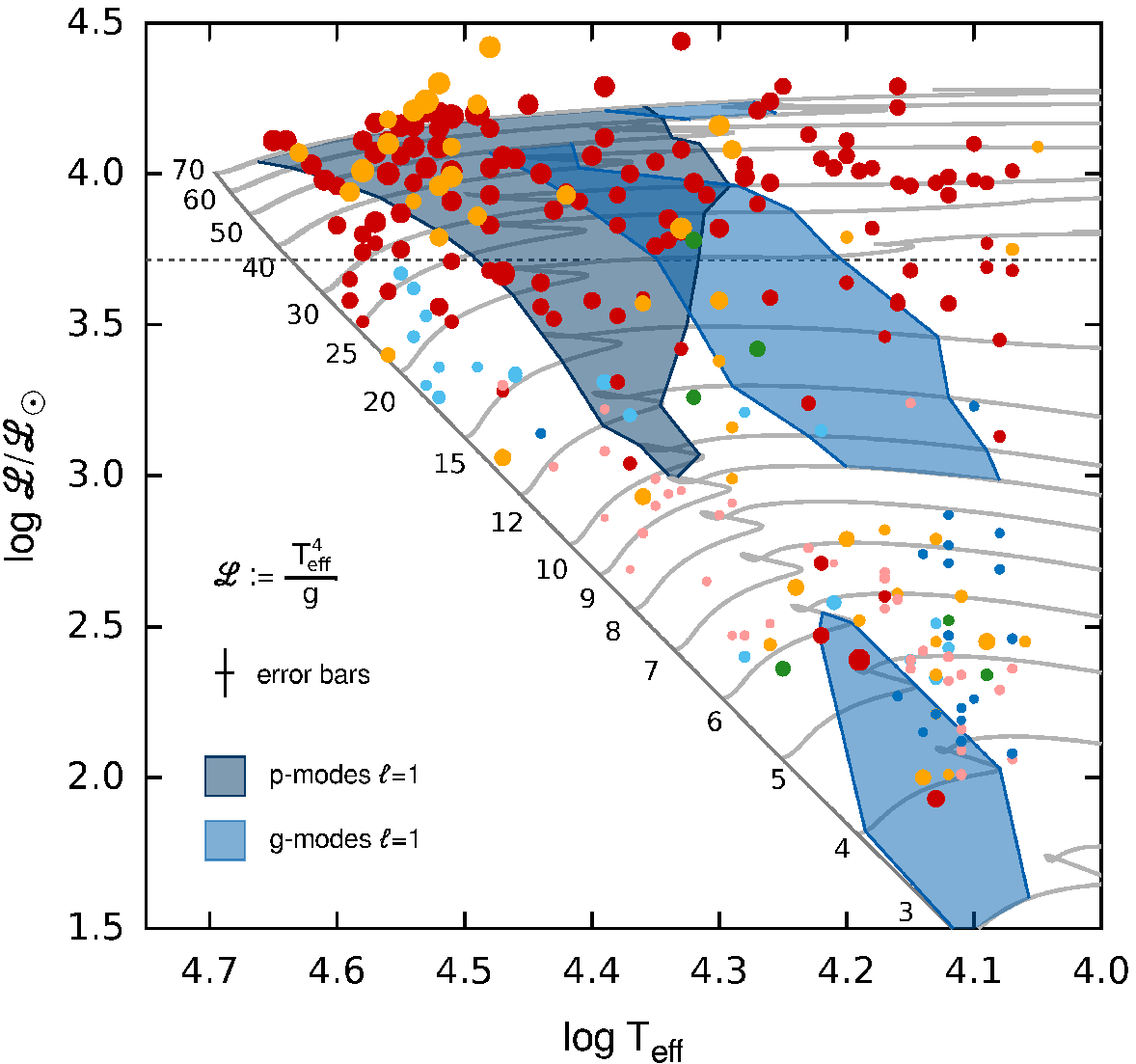}
\includegraphics[width=0.47\textwidth]{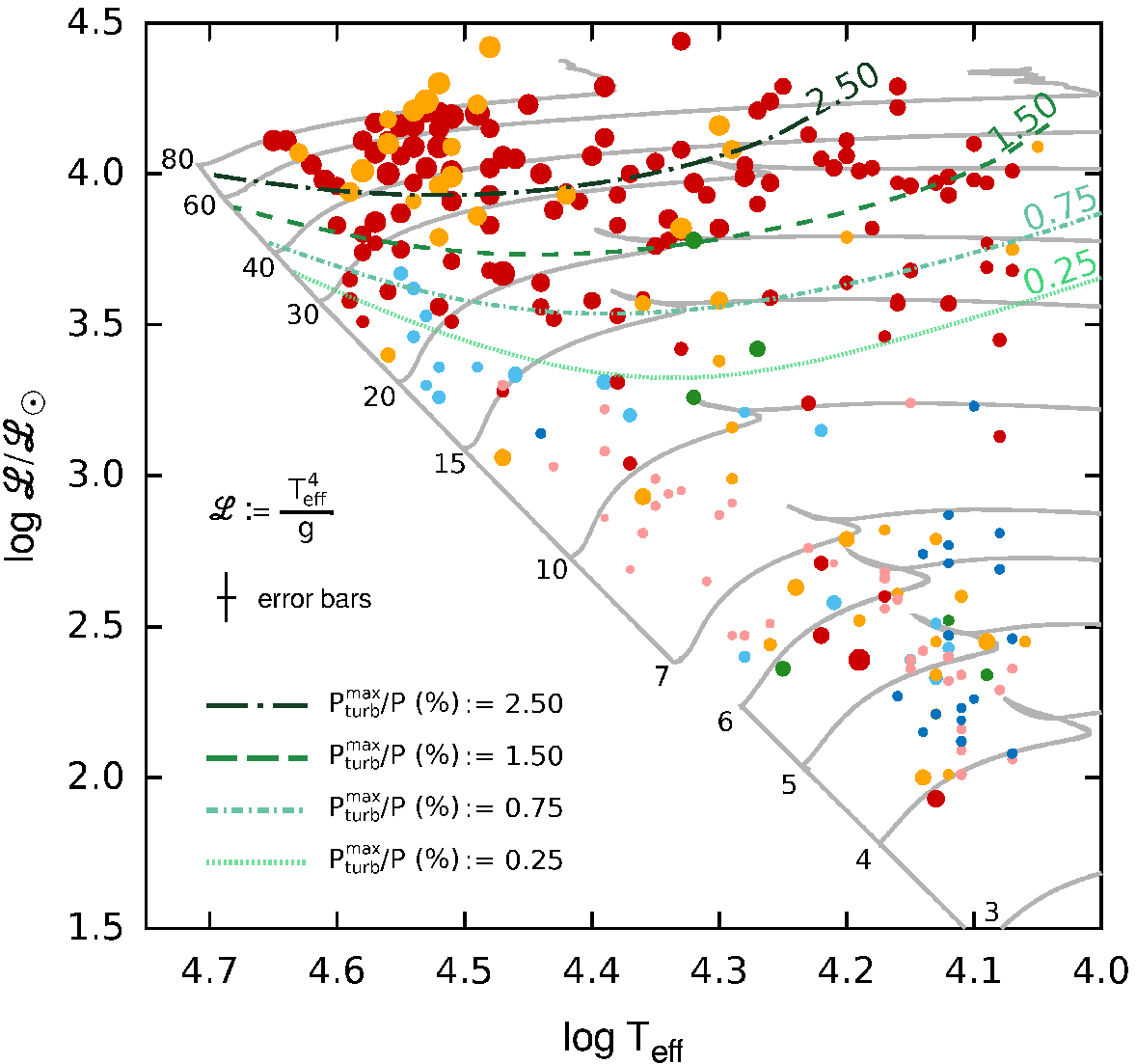}
\caption{Model predictions for [left] heat-driven p- and g-modes with $\ell$=1 computed by \cite{God16}, and [right] the maximum fraction of turbulent pressure over total pressure as derived by \cite{Gra15a}. 
Same observations as those included in right panel in Fig~\ref{fig-f5}, with the color code 
described in Sect.~\ref{sec-resul-qualit} and symbol sizes proportional to \vmac. We note that 
each panel consider different evolutionary tracks for coherence with the models considered in
the original papers performing the computations.}
\label{fig-f6}
\end{figure*}

\subsection{Using observations to evaluate the pulsational scenario of macroturbulent broadening in OB stars}\label{sec-resul-obs}

There are many aspects that must be considered to evaluate the suggested macroturbulent/pulsational broadening connection and, in the framework of this scenario, identify the excitation mechanism starting the instabilities which, ultimately lead to the extra-broadening of the line-profiles. In this context, it is important to remark that different types of oscillations, combined with other geometrical factors, can give raise to different spectroscopically observable features (mainly in the form of variable line-profile shape, asymmetry and broadening). We refer the reader to the works by \cite{Sch97} and \cite{Tel97} for a comprehensive illustration of predicted line-profile variations originated by adiabatic non-radial pulsations in rotating stars. Therefore, we ideally have to assess -- for the various proposed driving mechanisms -- how the predictions about all these (time-dependent) line-profile characteristics confront with the observations. In this regard, we basically enter in the domain of asteroseismology.

Time-resolved spectroscopy is the best suitable way to unravel whether the line-profile behavior is connected with strict periodicity as occurs for long-term coherent modes or rather with stochastic excitation. It is also a very powerful tool to evaluate whether the line-broadening and variability is connected with a few excited modes or is the result of collective effect of waves. Last, it can also help to identify whether there is a dominance of p- or g-modes acting on the line-profiles. However, following \cite{Sim12}, we also illustrate in next section how the empirical distribution of the single snapshot line-broadening properties in the sHR diagram discussed in Sect.~\ref{sec-resul-sHR} can be used as a complementary (and definitely {\em less expensive} compared to time-resolved spectroscopy) tool to assess the viability of the possible connection between macroturbulent broadening in OB stars and some of the excitation mechanisms of (possibly-)cyclic surface motions indicated above.

\subsubsection{Distribution of macroturbulent broadening in the sHRD and driving mechanisms of stellar oscillations}\label{sec-resul-obs-broad}

In \cite{Sim15a} we used part of the observational sample presented here to investigate possible correlations between the location of stars with a remarkable macroturbulent broadening contribution and the high-order g-mode instability strips from \cite{Miglio2007a} and \cite{Godart2011}. Later on, the same observations were included in the work by \cite{Gra15a} to assess the proposed connection between turbulent pressure fluctuations initiated in inefficient (in transporting energy) sub-surface convection zones and the occurrence of macroturbulent broadening in OB stars. 

In Fig.~\ref{fig-f6} we present again the comparison of observations and model predictions for both scenarios but, this time, using the complete, cleaned sample\footnote{For comparative purposes with the observational dataset used in \cite{Sim15a} and \cite{Gra15a}, the sample considered here includes $\approx$150 stars more and excludes $\approx$20 stars afterwards detected to be spectroscopic binaries using 
new multi-epoch observations compiled since then.} discussed in Sect.~\ref{sec-resul-sHR}.  Similarly to right panel in Fig.~\ref{fig-f5}, we combine all stars from groups A and B in the same plot, using the color code defined in Fig.~\ref{fig-f4}, symbol sizes proportional to \vmac, and excluding those stars for which this quantity could not properly determined (see Sect.~\ref{sec-resul-vsvm}).

The left panel of Fig.~\ref{fig-f6} is an updated version of Fig.~2 from \cite{Sim15a} in which we present results in the sHR diagram (instead of the \grav-log\Teff\ diagram), and include the instability strips for heat-driven p- and g-modes with $\ell$=1 resulting from a new homogeneous set of adiabatic and
non-adiabatic computations by Godart et al.\ (2016) in the whole 3\,--\,70~M$_{\odot}$ range\footnote{Although we concentrate here on $\ell$=1, Godart et al.\ (2016) actually include and discuss results for $\ell$=1\,--\,20.}. 
As already commented in \cite{Sim15a}, the presence of large red circles in the upper part of the sHR diagram outside the predicted instability domains implies a strong empirical challenge to non-radial modes excited by a heat mechanism being the main physical driver of the non-rotational broadening affecting O stars and B~Sgs \citep[except maybe for the early-B Sgs, located inside the post-TAMS g-mode strip for stars with $M\gtrsim$~10~\msun, see also notes in][]{Aer09}. Indeed, even only considering the stars located inside instability domains we would expect a different effect on the profile shape of the low degree p- and g-modes, the latter being characterized by having a denser frequency spectrum of excited modes. But this is not what the observations are telling us (see Sects.~\ref{sec-resul-sHR} and \ref{sec-resul-step-var}). 

However, this is not the last word concerning this scenario. As indicated above, the instability domains presented in the left panel of Fig.~\ref{fig-f6} correspond to computations for low degree modes. The situation may improve when accounting for the predictions for higher degree modes\footnote{We note that the simulations by \cite{Aer09} considered 241 excited modes with degree $\ell$ from 1 to 10.}. In addition, it is important to remark that the use of different input parameters\footnote{For reference, the instability domains presented in left panel of Fig.~\ref{fig-f6} correspond to computations performed with the ATON evolutionary code \citep{Ventura2008} assuming the OPAL opacity tables \citep{Iglesias1996}, a metallicity Z=0.015, and the metal mixture from \cite{Grevesse1993}. See \cite{God16} for further details.} (e.g. opacities, metal mixture, metallicity, overshooting parameter) or even evolutionary codes may alter the results \citep[see, e.g.,][for recent illustrations of these effects]{Miglio2007a, Miglio2007b, Zdravkov2008, Salmon2012, TurckChieze2013, Mar13, Cas14}.  

It will hence be interesting to investigate how all these effects may help to bring instability domains -- in particular, those connected with g-modes -- towards the ZAMS in the O/early-B star domain, and towards cooler temperatures in the B supergiant region, in a reasonable way. In this context, we already indicate, for example, that the instability strips recently computed by \cite{Mor16} based on the new Fe opacity measurements by \cite{Bailey2015} reach all the way to the ZAMS. Concerning the late-B~Sgs, oscillations excited by the $\epsilon$-mechanism \citep{Noe86, Unn89, Mor12b} may play an important role. Last, not to forget the predicted occurrence of adiabatic strange modes for log$\mathscr{L}/\mathscr{L}_{\odot}$~$>$~3.7 (these refer to modes for which the excitation in the iron opacity bump due to the $\kappa$-mechanism is enhanced by the large amplitude of the mode trapped into a superficial cavity; see, e.g, \citeauthor{Saio1998}~\citeyear{Saio1998}, and references therein; \citeauthor{Aer10a}~\citeyear{Aer10a}). Some of these points are explored and discussed in more detail in the parallel work by \cite{God16}. 

The right panel of Fig.~\ref{fig-f6} is an adaptation of Fig.~4 from \cite{Gra15a} in which we over-plot the isocontours of the quantity P$_{\rm turb}$(max)/P (maximum fraction of turbulent pressure over total pressure) as predicted by \cite{Gra15a} to the distribution of line-broadening properties of the updated sample of stars in the sHR diagram. For completeness, the information presented in this panel must be complemented with Fig.~5 in \cite{Gra15a} in which \vmac\ is plotted against P$_{\rm turb}$(max)/P. 

As already pointed out by \citeauthor{Gra15a}, the good correlation found between these two quantities render the turbulent pressure scenario as a very promising one, at least regarding the upper part of the HR diagram. Indeed, the predicted behavior of the strength of this driving mechanism with \Teff\ and $\mathscr{L}$ may explain the similarity of line profiles found in the whole O and B~Sgs domain -- this scenario implies only one
broadening agent behaving in a similar way in the whole range -- and the strong dependence of \vmac\ with $\mathscr{L}$ established empirically (see Sects.~\ref{sec-resul-sHR}). However, there is still a very important point to be evaluated in this scenario. \citeauthor{Gra15a}, only provided indirect arguments indicating that turbulent pressure fluctuations initiated in sub-surface convection zones may drive high-degree 
oscillations. One could extend further this argument, indicating that we might be dealing with stochastically excited waves behaving in a similar way as the (gravity) modes assumed in the simulations by \cite{Aer09}. However, all this line of argument must still be directly proven, especially regarding the fact that we must end up reproducing the observed amount of line-broadening and profile shapes characterized by not having  any detectable substructure in the spectral lines. The latter is indeed an important observational constraint to be taken into account when assessing any scenario to explain macroturbulent broadening in OB stars.

Some work in this line has been recently performed by \cite{Aer15}, who also proposed a third (maybe complementary) scenario in which macroturbulent broadening is related to the occurrence of convectively driven internal gravity waves (IGWs). \cite{Aer15} based their study in two-dimensional (2D) numerical simulations as in \cite{Rog13} and showed that the stochastic spectrum of running waves caused by IGWs may lead to detectable line-profile variability and explain the extra-broadening observed in the line-profiles of OB stars. However, given the intense CPU time required by the computations they could only perform the investigation for a 3~\msun\ main sequence star.  Therefore, we still can not compare the predictions of this scenario with the empirical distribution of line-broadening properties in the sHR diagram as in previous cases. 

New simulations of such internal gravity waves across stellar evolution is hence a very interesting line of future work in this respect. In addition, as indicated in Sect.~\ref{sec-resul-obs}, and illustrated in next section, the compilation and analysis of spectroscopic time-series will also provide interesting clues to identify the type of waves and the associated driving mechanism producing the different sources of non-rotational broadening and line-profile variability detected in O stars and B~Sgs (as asteroseismology has efficiently shown in the case of B stars -- among many other type of stars across the HR diagram).

\subsubsection{Macroturbulent broadening and line-profile asymmetry}\label{sec-resul-obs-asym}

Apart from broadening effects, line asymmetries and shifts are important observables for the characterization of the seismic properties of stars, in particular if we include a description of how these line properties vary with time \citep[cf. Chapters 5 and 6 of][]{Aer10b}. It is hence of interest to incorporate quantities accounting for this information to achieve any physical interpretation of macroturbulent broadening, especially in the context of stellar oscillations. 

The velocity moments of the line-profile \citep{Bal86, Aer92, Sch97} have been proven to be powerful tools to characterize the temporal behavior of line asymmetry independently of the physical cause that is originating it. An interesting work in this respect is the one by \cite{Sch97}, where line-profile asymmetry due to a long lifetime oscillation mode of degree $\ell$ and azimuthal order $m$ in rotating stars is shown for a large variety of oscillation and rotational parameters (the latter quantified in terms of the rotational frequency $\Omega$ and the inclination angle $i$ of the star). The predicted moment variations along with the line-profiles are presented for many combinations of $(\ell,m,i,\Omega)$, showing that asymmetry is obvious from the moments even though the profiles themselves are seemingly symmetric. In the same line, \cite{Bri04} demonstrated that the moments of a line-profile allow to distinguish line asymmetry due to surface spots versus oscillations and \cite{Hek10} showed the utility of using the moments even in the case of stochastically excited oscillations. While these are just three illustrative examples, many other works in the last decades have shown the interest of quantifying line-profile variations by means of the line moments.

\begin{figure}[!t]
\centering
\includegraphics[width=0.48\textwidth, angle=0]{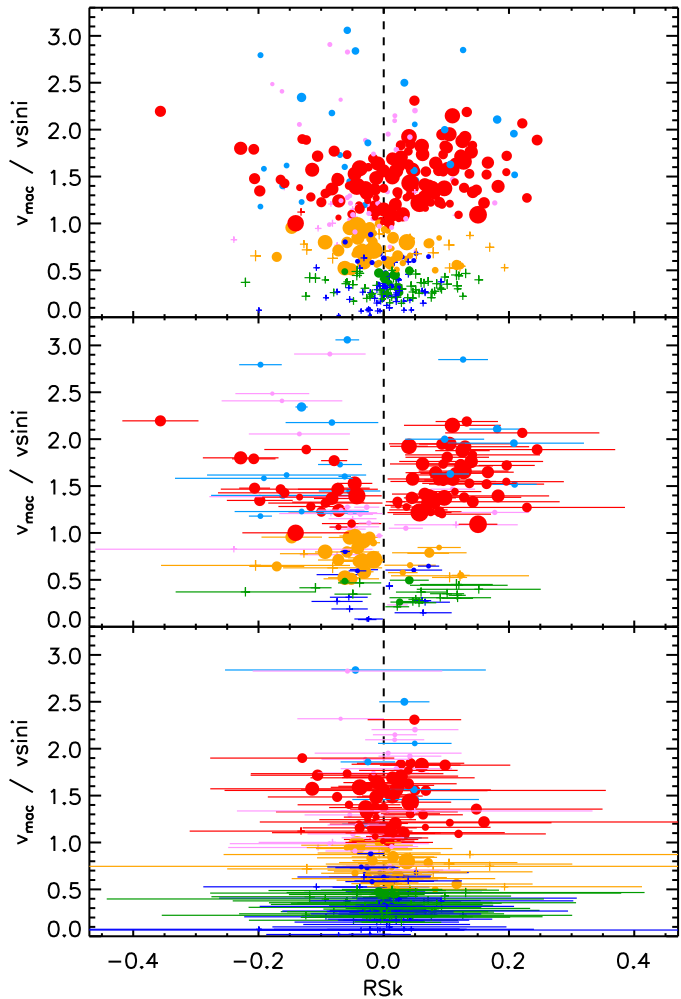}
\caption{Macroturbulent broadening (normalized to \vsini) versus asymmetry for [Top] all 432 stars considered in this paper -- no uncertainties overplotted --; [Middle] only those stars with clear asymmetric profiles; [Bottom] stars with line-profiles in which the measured value of the quantity $RSk$ is compatible with a symmetric profile given the associated uncertainty. Same colors and symbols as in Figures~\ref{fig-f4} and \ref{fig-f5}. The size of the symbols scales with \vmac.}
\label{fig-f7}
\end{figure}

In this section, we use our single-epoch observational dataset to investigate possible correlations between \vmac\ and the amount of asymmetry of the line-profiles. Being aware of the limitations of this study -- given the expected variability of the considered quantities with time \cite[see, e.g.,][]{Sim10, Aer14} -- we still consider of interest the discussion of the results of such a single snapshot approach while awaiting for the complete on-going time-resolved IACOB observations. To this aim we plot in Fig.~\ref{fig-f7} the dimensionless ratio \vmac/\vsini\ versus the, also dimensionless, quantity $RSk$ ({\em relative skewness}) defined in Sect.~\ref{sec-asym-tools}. For a cleaner presentation of results, we divide the figure in three panels. From top to bottom we include (a) the whole sample of 432 stars discussed in Sect.~\ref{sec-resul-quantit} without uncertainties in $RSk$ over~plotted, (b) only those stars with clear asymmetric profiles, and (c) the remaining stars, in which the measured $RSk\pm\sigma_{RSk}$ is compatible with a symmetric profile\footnote{We remind that, formally, a perfectly symmetric line implies $RSk$\,=\,0, while positive/negative values of the quantity $RSk$ represent a redwards/bluewards asymmetric line (see, e.g., panels A.3 and A.1 in Fig.~\ref{fig-f3}, respectively).}. 

Figure~\ref{fig-f7} is complemented with the information provided in Table~\ref{tab-tx}, where we indicate the percentage of stars -- in each of the subgroups defined in Fig.~\ref{fig-f2} -- having line-profiles with clear positive/negative {\em relative skewness} or being compatible with $RSk$\,=\,0.

\begin{table} [t!]
\small
\begin{center}
\caption{Summary of results presented in Fig.~\ref{fig-f7} regarding the quantity $RSk$. Asymmetric/symmetric profiles refer to cases with |$RSk$| -- $\sigma_{RSk}$ greater/less than zero, respectively. Stars separated by subgroups as defined in Sect.~\ref{sec-resul-vsvm} (see also Fig.~\ref{fig-f2}).}
\label{tab-tx}
\begin{tabular}{llccccc}
\hline\hline
\noalign{\smallskip}
              &       &    &    \multicolumn{2}{c}{Asymm.} & & Symm.  \\
\cline{4-5} \cline{7-7}
\noalign{\smallskip}
Color         & Label &\#    &    $RSk <$~0 & $RSk >$~0 & &  \\
\hline
\noalign{\smallskip}
Red    & A.1 & 137 & 17\% & 41\% & & 42\%  \\
Orange & A.2 & 70  & 33\% & 12\% & & 55\%  \\
Cyan   & A.3 & 27  & 48\% & 22\% & & 30\%  \\
Green  & B.1 & 73  &  7\% & 19\% & & 74\%  \\
Pink   & B.2 & 49  & 39\% & 10\% & & 51\%  \\
Blue   & B.3 & 76  & 12\% &  8\% & & 80\%  \\
\noalign{\smallskip}
Total  & ---    & 432 & 21\% & 22\% & & 57\% \\
\noalign{\smallskip}
\hline
\end{tabular}
\end{center}
\end{table}

The first interesting result to be highlighted is the presence of a remarkable number of stars with clearly identified asymmetric profiles (almost half of the global sample, see last row in Table~\ref{tab-tx}). Among them, the percentage of profiles with positive/negative skewness is basically the same ($\approx$22\%). This result can be interpreted as empirical evidence of line asymmetries being associated with a variable phenomenon which is producing a time-dependent distribution of motions in the line formation region with an average velocity close to zero. Since we are considering snapshot measurements for a large number of stars, and each of these measurements can be randomly picked at any instance during the variation cycle, such a phenomenon would result in a distribution of points similar to that shown in Fig.~\ref{fig-f7}. 

This would exclude, e.g., any wind-type variability or differential rotation as main cause of the asymmetries -- at least generally speaking --, as these two cases imply motions that are going along with an ``average'' velocity different from zero (i.e. is very difficult to obtain line-asymmetries which swap sign). In contrast, stellar oscillations fulfill this requirement. 

Table~\ref{tab-tx} also shows that, except for the cases in which the relative contribution of the macroturbulent broadening is negligible (i.e. B.1 and B.3, where \vmac~/\vsini\ $<$~0.5), the percentage of stars with clear asymmetric profiles is always larger than 45\%. In addition, back in Fig.~\ref{fig-f7}, the larger concentration of profiles found among the most asymmetric lines are those having a dominant contribution of the macroturbulent broadening (red dots). All this suggests the potential connection between macroturbulent broadening and line-asymmetries (and hence, following the argument above, a variable phenomenon).

As indicated by \cite{Aer09, Aer14} and \cite{Sim10}, the association between stellar oscillations, the detected asymmetries and the so-called macroturbulent broadening is a promising possibility. This scenario seems to be also supported, at least in a global sense, by the distribution of stars in the \vmac/\vsini\ vs. $RSk$ diagram presented in Fig.~\ref{fig-f7}. However, a closer inspection of the distribution of points in each of the subgroups considered in Table~\ref{tab-tx} indicates that the situation may be more complex than just assuming only one type of physical driving of the (likely variable) line-broadening and line-asymmetry in the whole O and B-type star domain. For example, the fact that the $RSk$ distribution in the middle panel of Fig.~\ref{fig-f7} is weighted to positive values in group A.1 (red) and negative values in groups A.2 (orange), A.3 (cyan) and  B.2 (pink) -- see also Table~\ref{tab-tx} -- deserves further attention in future studies incorporating time-resolved information about all the considered quantities. 

One interesting possibility to be investigated would be related to the different effect that pressure and gravity-dominated oscillations (which have a dominant radial and transversal component, respectively) may have on snapshot properties of the line-profiles in a large sample of stars, since their amplitude distribution across the oscillation cycle in terms of asymmetry is different \citep[e.g., Chapter 6 in][]{Aer10b}. Other options to be evaluated would be the effect that stellar winds and magnetic fields may have on the skewness properties of the considered diagnostic lines. 

The analysis of time-series observations of a large sample of stars covering the various types of line-profiles will allow to investigate whether the observed distribution of points in Fig.~\ref{fig-f7} is connected to the type of oscillations indicated above. In addition, the incorporation of information about the stellar wind and magnetic properties of the considered stars -- along with the single/composite status of the analyzed spectra -- will serve to identify those cases in which the shape and variability of the  considered diagnostic lines can be affected by other effects apart from stellar oscillations.

\subsubsection{Macroturbulent broadening and line-profile variability: OB stars vs. well known pulsating and spotted B stars}\label{sec-resul-step-var}

\begin{table*} [t!]
\small
\begin{center}
\caption{Results from the time-dependent line-broadening analysis of a sample of eight O and B stars with an important contribution of non-rotational sources of line-broadening.}
\label{tab-t2}
\begin{tabular}{llcccccccccccl}
\hline\hline
\noalign{\smallskip}
\# & Star    &    Sp.C & N &  \multicolumn{3}{c}{\vsini} & & \multicolumn{3}{c}{\vmac} & log T$_{\rm eff}$ & log $\mathscr{L}/\mathscr{L}_{\odot}$ & Notes \\
\cline{5-7} \cline{9-11}
\noalign{\smallskip}
&  &                  &  & $\bar{x}$  & $\sigma_{\rm x}$ & Range & & $\bar{x}$  & $\sigma_{\rm x}$ & Range & & & \\
\hline
\noalign{\smallskip}
\multicolumn{12}{l}{O stars and B~Sgs (OB stars) with multi-epoch observations gathered in the framework of the IACOB project} \\
\hline
\noalign{\smallskip}
1 &   HD~199579 & O6.5~V((f))z & 112 &  56.2  &  8\% &  [48.0, 69.0]   &  &  81.2  &  7\% &  [65.0, 90.4] & 4.60  &  3.83  & SB1\\
2 &   HD~36861  & O8~III       & 172 &  54.2  &  3\% &  [49.4, 60.0]  &  &  73.2  &  4\% &  [64.3, 80.0]  & 4.55  &  4.06  & -    \\
3 &   HD~38771  & B0~Iab       & 154 &  61.5  &  7\% &  [49.6, 71.4]  &  &  82.5  &  6\% &  [67.5, 95.2] & 4.47  &  4.06  & -  \\
4 &   HD~34085  & B8~Iab       & 108 &  38.7  &  7\% &  [32.3, 47.6]  &  &  52.9  &  10\% &  [47.1, 67.5] & 4.10  &  4.10  & -   \\
\noalign{\smallskip}
\hline
\noalign{\smallskip}
\multicolumn{12}{l}{B main sequence stars from the study by \cite{Aer14}} \\
\hline
\noalign{\smallskip}
5 &  HD~111123  & B0.5~IV      & 119 &  28.6  &  22\% &  [16.5, 41.9]  &  &  42.8  &  11\% &  [33.1, 57.3] & 4.43  &  3.42  & SB1, nr-p  \\
6 &  HD~16582   & B2~IV        &  59 &  11.6  &  23\% &  [4.1, 16.1]   &  &  17.7  &  10\% &  [12.7, 20.8] & 4.36  &  3.04  &  r   \\
7 &  HD~105382  & B6~III       & 106 &  68.3  &  10\% &  [49.6, 78.8]  &  &  27.6  &  40\% &  [10.8, 73.0] & 4.24  &  2.15  & spot \\
8 &  HD~181558  & B5~III       &  30 &  4.1   &  100\% &   [0, 24.7]    &  & 15.4   &  48\% &  [ 0.0, 27.8] & 4.17  &  1.86  & nr-g   \\
\hline
\end{tabular}
\end{center}
\end{table*}

With the aim to support the pulsational hypothesis to explain macroturbulent broadening in massive O and B stars and validating state-of-the-art methodologies to estimate projected rotational velocities in these stellar objects, \cite{Aer14} investigated a sample of well known spotted and pulsating B stars with time-dependent profiles. They analyzed available multi-epoch spectroscopic observations by means of similar techniques as those described in \cite{Sim14} and used in this paper. They found that both line-broadening parameters vary (in anti~phase with each other) appreciably during the pulsation cycle whenever the pulsational and rotational velocity fields have similar
magnitudes. In addition, the macroturbulent velocities derived while ignoring the pulsations can vary by tens of \kms\ during the pulsation cycle. This study hence warned about misinterpretations based on a \vsini+\vmac\ single snapshot approach for the line-broadening characterization of stars with an important non-rotational broadening contribution to the line-profiles. 

Here we use already available multi-epoch observations to tackle the question how this situation compares with the case of OB stars with line profiles dominated by macroturbulent broadening and provide some first results using time-resolved spectroscopy to assess the pulsational hypothesis in these more massive objects without awaiting the full time series data.

We selected for this study four representative examples among the bright O stars and B~Sgs for which we have been gathering spectroscopic time-series during the last 4-6 years as part of the IACOB project. We complement this sample of stars with another four representative 
targets selected from the study by \cite{Aer14} for purposes of comparison. A total of $\approx$\,30\,--\,170 high-resolution spectra per target were analyzed in a homogeneous way, extracting information about \vsini, \vmac, and $RSk$ from each of the considered line-profiles (\ion{O}{iii}~$\lambda$5591, \ion{Si}{iii}~$\lambda$4552, \ion{Mg}{ii}~$\lambda$4481, or \ion{Si}{ii}~$\lambda$4128, depending on the spectral type). 

\begin{figure*}[!t]
\centering
\includegraphics[height=0.98\textwidth,angle=90]{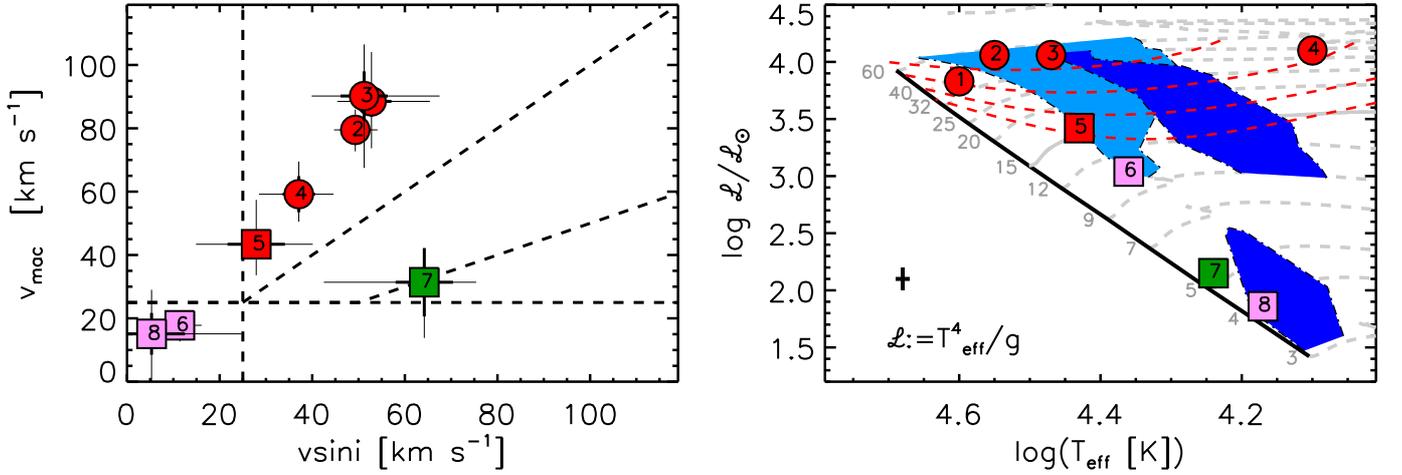}
\caption{Location of the eight stars discussed in Sect.~\ref{sec-resul-step-var} in the \vsini\,--\,\vmac, and sHR diagrams. Same color code as is previous figures. The four pulsating/spotted B main sequence stars considered by \cite{Aer14} are identified with square symbols, while the remaining OB stars surveyed by the IACOB project, with still unknown pulsational properties are marked with circles. Cyan and dark blue shaded region in the sHR diagram indicate the instability domains for p- and g- heat-driven modes, respectively, with $\ell$=1 computed by \cite{God16}. Dashed red lines in the same figure indicates the isocontours of the quantity P$_{\rm turb}$(max)/P as computed by \cite{Gra15a}.}
\label{fig-f11}
\end{figure*}

The list of targets considered for this study is presented in Table~\ref{tab-t2}, where we also summarize the results from the time-dependent line-broadening analysis\footnote{We remark again the homogeneity of the line-broadening analysis for the eight stars, which was performed by the same person (SS-D), with the same tool ({\sc iacob-broad}) and making the same assumptions.}, the derived stellar parameters (from the best SNR spectrum)\footnote{In the case of the four stars presented in \cite{Aer14}, we assumed the \Teff\ and \grav\ values provided in their Table~1.} and provide some notes about the type of variability.

The eight stars are located in the \vsini\,-\,\vmac\ and sHR diagrams in Fig.~\ref{fig-f11}. The interest and characteristics of these diagrams have been already described in previous sections. In this case, we incorporate to the \vsini\,-\,\vmac\ diagram information about the standard deviation and scatter resulting from the \vsini\ and \vmac\ measurements of the complete time-series (thick and thin horizontal/vertical lines, respectively). We also overplot in the sHR diagram the instability domains for p- and g- heat-driven modes with $\ell$=1 (cyan and dark blue shaped region, respectively) obtained by \cite{God16}, and the isocontours (red dashed lines) of the quantity P$_{\rm turb}$(max)/P computed by \cite{Gra15a} as described in Sect.~\ref{sec-resul-obs-broad}. 

Figure~\ref{fig-f12} illustrates the type of line-profile shape and variability associated with each star. In particular, each panel in this figure includes three profiles, corresponding to different epochs in which the line-asymmetry analysis has resulted in the maximum, minimum, and mean values of $RSk$, respectively, as derived from the whole time-series for a given star. This figure is complemented with Fig.~\ref{fig-f13}, which shows the complete distribution of points in the $RSk$\,--\,\vsini/\vmac\ diagram resulting from the line-broadening and line-asymmetry analysis of the full spectroscopic dataset per target.\\

We highlight below some interesting results from this study:

%
\begin{figure*}[!t]
\centering
\includegraphics[height=0.98\textwidth,angle=90]{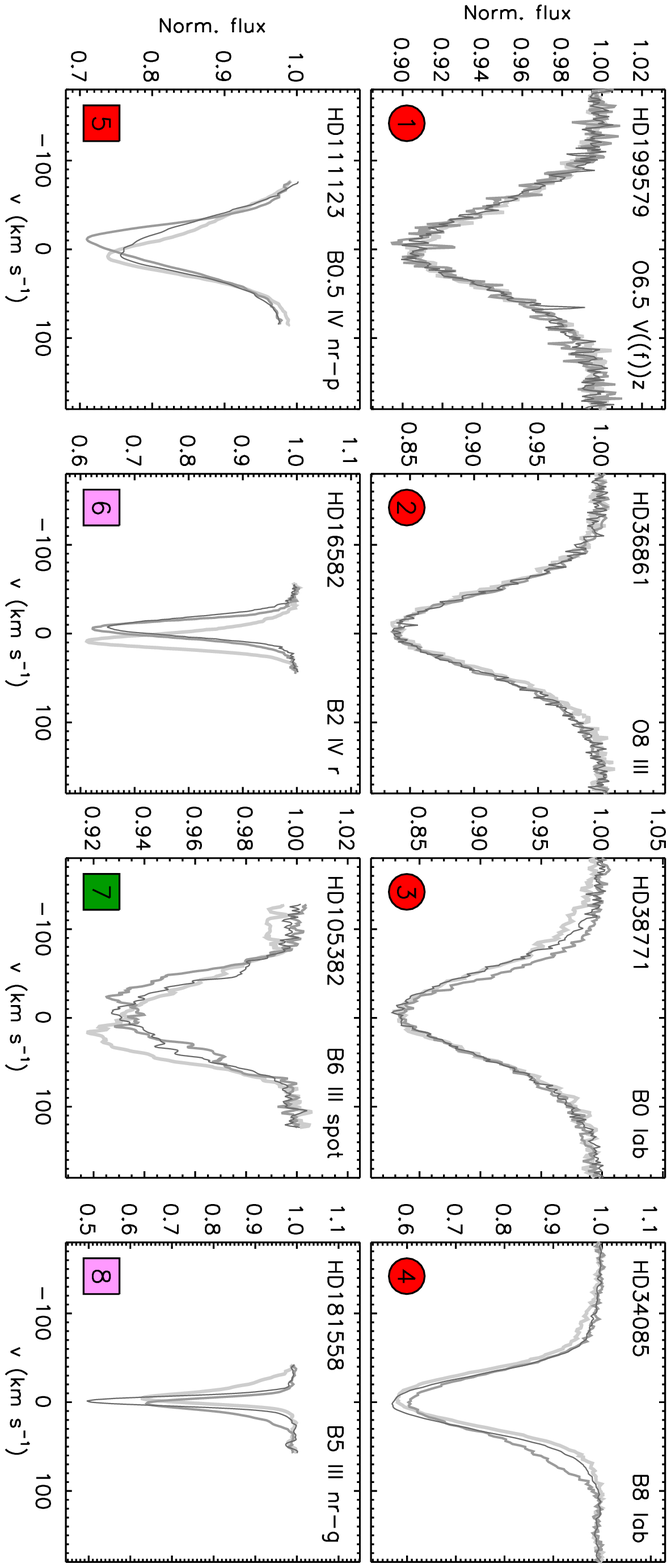} 
\caption{Examples of line-profile shape and variability for the sample of stars quoted in Table~\ref{tab-t2} and discussed in Sect.~\ref{sec-resul}. The three profiles shown in each panel correspond to different epochs in which the line-asymmetry analysis has resulted in the maximum, minimum, and mean values of $RSk$, respectively, as derived from the whole time-series (see gray filled circles in Fig.~\ref{fig-f13}). Note: HD~199579 and HD~111123 are two single line spectroscopic binaries; in both cases the line-profiles have been shifted in radial velocity to account for the orbital motion of the stars.}
\label{fig-f12}
\end{figure*}
%

%
\begin{figure*}[!t]
\centering
\includegraphics[height=0.98\textwidth,angle=90]{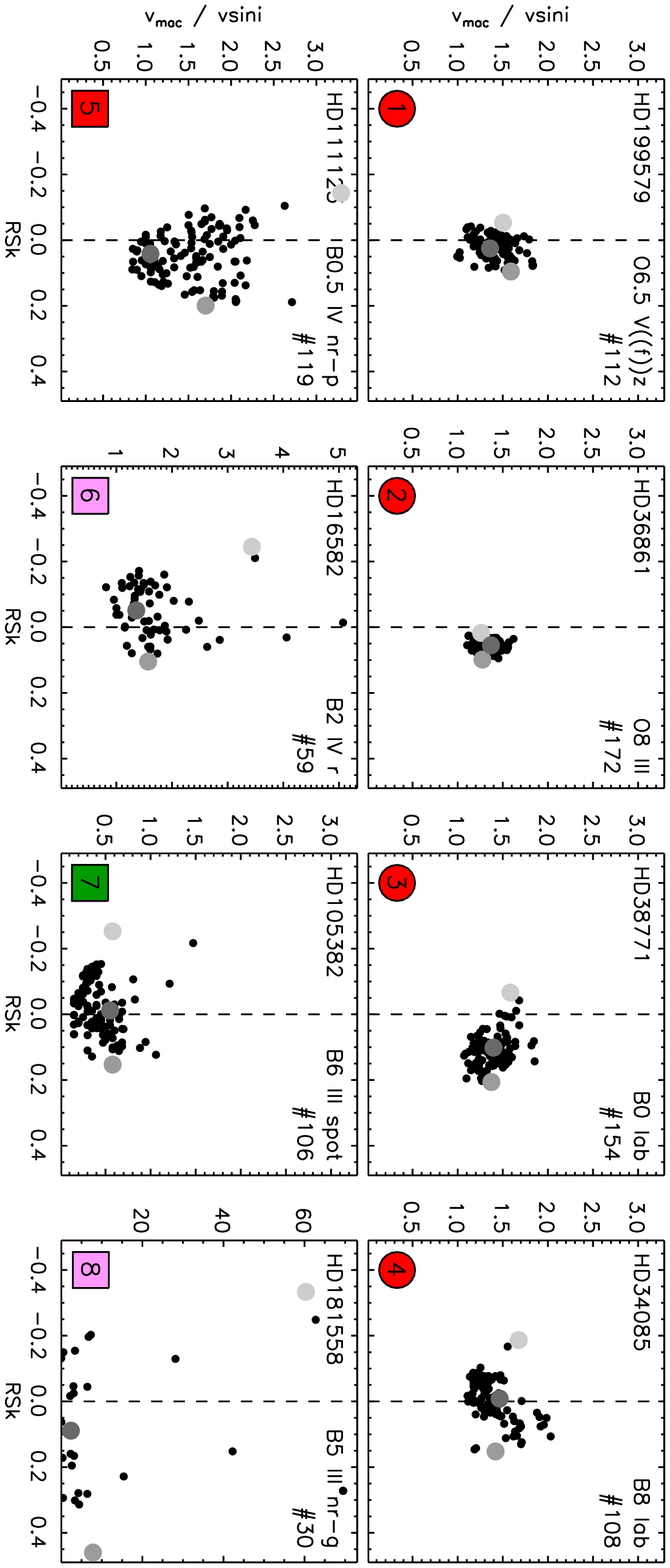} 
\caption{Distribution of points in the $RSk$\,--\,\vsini/\vmac\ diagram resulting from the line-broadening and line-asymmetry analysis of the full time-series for the sample of stars quoted in Table~\ref{tab-t2}. The three gray circles highlighted in each panel correspond to the line-profiles shown in Fig.~\ref{fig-f12} (i.e., those profiles per target having minimum, maximum and close to the mean values of the quantity $RSk$). We note the different scale in the y-axis in each panel.
}
\label{fig-f13}
\end{figure*}
%

\begin{itemize}
\item First, we remark that in O stars and B Sgs the scatter of the \vsini\ and \vmac\ measurements resulting from the line-broadening analysis of the spectroscopic time-series is not as extreme as in the pulsating/spotted B main sequence stars (see Table~\ref{tab-t2}). In particular, the standard deviation in the derived \vsini\ is always less than 10\% in stars \#1--4, to be compared with the much larger values obtained for stars \#5--8. Therefore, the warning about the reliability of \vsini\ measurements highlighted by \cite{Aer14} is less critical for luminous OB stars, even though their line-profiles have an important contribution of the macroturbulent broadening and show variability and asymmetries.
\item A direct (qualitative) inspection of Fig.~\ref{fig-f12} already shows that while in all cases, sources of non-rotational broadening are clearly shaping the line-profiles, the global shape and type of line-variability is significantly different in both samples. Indeed, the occurrence of line-profile variability is most clearly detected by-eye in the pulsating/spotted B main sequence stars and the considered B~Sgs than in the two investigated O-type stars. In addition, the amount of variability is much larger in the B star sample, and the type of variability is qualitatively different.
\item There is a variety of shapes and types of line-profile variability in the B sample; in contrast, these two characteristics are more homogeneous in the OB sample (even though, as illustrated in the right panel of Fig.~\ref{fig-f11}, this latter group of stars cover a wide range in evolutionary status and stellar properties).
\item The amount of extra-broadening is much larger in the O stars and B Sgs than in the pulsating or spotted B main sequence stars (see Table~\ref{tab-t2} and left panel in Fig.~\ref{fig-f11}).
\item The spotted B star (HD~105382, \#7) is clearly separated in the \vmac/\vsini\ diagram (Fig.~\ref{fig-f11}, left panel) from the rest of the considered targets. As indicated by \cite{Aer14} the spectroscopic variability produced by the spot can be mimicked in terms of a variable combination of \vsini\ and \vmac; however, the relative contribution of the non-rotational broadening is considerably smaller in this star (even the derived \vmac\ ranges between $\approx$20 and 40~\kms). 
\item Among the pulsating B main sequence stars, the one having a larger \vmac\ and a profile-shape closer to those found in the other group of stars is the low-amplitude multi~periodic non-radial pressure-mode pulsator HD~111123 (\#5, Aerts et~al. 1998; Cuypers et~al.  2002). However, the type of variability detected in this star is different from the one observed in the other considered O stars and B~Sgs. In particular, the variability is concentrated in the core of the line in the case of HD~111123, while in the OB stars the larger variability is detected in the wings. 
\item While the B main sequence star identified as a non-radial gravity mode pulsator \citep[HD~181558, \#8][]{Dec02} is one of the candidates to show an important macroturbulent broadening contribution, the global broadening of the associated line-profiles is much smaller compared to what is found in OB stars. We note, however, that this characteristic may be explained by the fact that the star present one dominant g-mode while the collective effect of many modes is required to result in a much broadened line-profile. 

\item There is a fairly good qualitative agreement between the predicted pulsational properties of the stars in terms of heat-driven modes -- as indicated by the instability domains --, and the observed line-profile variability\footnote{We note that even though HD~34085 (Rigel) is located far outside the considered instability domains, the variability detected in this star has been proposed to be produced by the $\epsilon$-mechanism \citep{Mor12a, Mor12b}.} (see middle panel in Fig.~\ref{fig-f11} and Fig.~\ref{fig-f12}, respectively). However, the connection between the global shape of the line-profiles and its location inside/outside instability domains is less clear, or even non-existent. Contrarily, as indicated by \citet[][see also Sect.~\ref{sec-resul-obs-broad}]{Gra15a} there seems to be a better correlation between the amount of extra-broadening and the quantity P$_{\rm turb}$(max)/P.

\item All stars, except HD~199579 (\#1) and HD~36861 (\#2) show a considerable dispersion in the quantity $RSk$ (see Fig.~\ref{fig-f13}). In the case of the pulsating/spotted B main sequence stars, the distribution of points around $RSk$\,=\,0 is roughly symmetric. The same occurs for HD~34085 (\#4). For HD~199579 (\#1), the distribution of points in the $RSk$\,--\,\vsini/\vmac\ diagram is compatible with a set of noisy (almost) symmetric line-profiles (see also Fig.~\ref{fig-f12}). Last, most of the analyzed line-profiles in the spectroscopic time-series of HD~36861 (\#2) and HD~38771 (\#3) have positive skewness.

\end{itemize}

This first combined investigation of line-broadening and line-profile variability in a small sample of O- and B-type stars highlights the need to incorporate time-series spectroscopy to study macroturbulent broadening, opening new sources of information for the understanding of this spectroscopic phenomenon \citep[as already proposed by][]{Aer14}. The results highlighted above seems to go in the same direction as the one indicated by the distribution of single snapshot line-broadening properties in the sHR diagram presented in Sect.~\ref{sec-resul-sHR} (see Fig.~\ref{fig-f5}). Namely, a different origin of the extra source of non-rotational broadening in stars with masses above $\approx$~15~\msun\ and B main sequence stars. In addition, the inclusion of multi-epoch information seems to indicate that there is not necessarily a connection between the observed line-profile variability and the amount of line-broadening.

\subsubsection{Concluding remarks}

In view of all the empirical clues presented in this paper, and the various scenarios proposed to explain the occurrence of macroturbulent broadening in the OB star domain (see Sect.~\ref{sec-resul-obs-broad}), we propose that the main observed properties of the line-profiles
of stars in the uppermost part of the sHR diagram are a result of the combined effect of (a) pulsation modes associated with a heat-driven mechanism and (b) possibly-cyclic surface motions initiated by turbulent pressure instabilities. 

Under this assumption, and combining the proposals by \cite{Aer09} and \cite{Gra15a}, the latter mechanism would be the main responsible of the large amount of non-rotational broadening detected in O stars and B~Sgs -- though further theoretical confirmation that the turbulent pressure scenario may end up in the observed broadening is still required --. Then, operating at the same time in certain regions of the sHR diagram, long-lived p- and g-modes excited by the $\kappa$- or $\epsilon$-mechanisms (plus maybe other spectroscopic variability agents, such as spots, variable winds, or strange modes) would be the main agents producing the observed line-profile variability (and hence asymmetries).

As a consequence, above a certain $L$/$M$ ratio (or equivalently, a certain value of log~$\mathscr{L}$), stars with line-profiles affected by a similar (dominating) macroturbulent broadening may present different type of line-profile variability depending on their location in the sHR diagram (as, e.g., predicted by the corresponding instability domains). Then, as the effect of turbulent pressure diminish when moving down towards the B star domain the observed line-profile characteristics are mainly produced by the effect of coherent heat-driven modes (or spots in certain cases). 

This combined scenario could easily explain why, e.g, HD~199579 (\#1) and HD~38771 (\#3) have very similar profiles in terms of global shape, but different type of variability. Indeed, no clear variability is detected in HD~199579, as expected from the predicted absence of excited heat-driven modes for O stars close to the ZAMS, while the variability found in HD~38771 could be explained in terms of the predicted gravity-mode oscillations with dominant tangential component. It would also explain why the profile of HD~111123 (\#5) is closer in shape to the line-profiles commonly found in O stars and B~Sgs, because the star is located in a region of the sHR diagram where the effect of turbulent pressure begins to be important (see middle panel in Fig.~\ref{fig-f11}). In addition, its type of line-profile variability is the one expected for a pressure-mode pulsator.  Similar arguments could be applied to the rest of stars in the sample discussed in this section. For example, the remarkable line-profile variability (mainly concentrated in the wings, as expected from a g-mode pulsator), but low value of \vmac, associated with HD~181558 (\#8). 

This is, hence, a scenario which allows to explain most of the empirical evidence presented along this paper, and which must be further evaluated using time-dependent spectroscopy of a larger sample of targets (from the observational point of view), along with other theoretical simulations in the line of those presented in \cite{Aer15}.

\section{Summary and future prospects}\label{sec-summary}

In this paper we provide new observational clues for the understanding of the empirical characteristics and physical origin of the so-called macroturbulent broadening in massive O and B stars. We base this study in a large, high-quality spectroscopic dataset compiled in the framework of the IACOB project during the last 7 years. This observational material comprises high-resolution ($R$\,=\,25000\,--\,85000), high signal-to-noise ratio (SNR$>$150) spectra of a sample of 432 Galactic blue massive stars with spectral types in the range O4\,--\,B9 and covering all luminosity classes. In other words, we consider stars born with $M_{\rm ZAMS}$ in the range $\approx$~4\,--\,80~$M_{\odot}$, and their evolved descendants up to effective temperatures of the order of 10000~K. The sample can be roughly divided in O stars and B~Sgs (i.e. OB stars) -- populating the uppermost part of the HR diagram --, and B dwarfs and giants -- located in or close to the main sequence phase --. We mainly concentrate in single snapshot spectra, but also consider available multi-epoch observations to identify and exclude from the studied sample clear spectroscopic binaries and to provide some empirical clues about the potential connection between macroturbulent broadening and line-profile variability.

We use a set of modern semi-automatic tools developed to perform quantitative spectroscopic analysis of large samples of O- and B-type stars in an objective, fast and reliable way. In this paper we concentrate on the information about the line-broadening (\vsini\ and \vmac) and spectroscopic (mainly \Teff\ and \grav) parameters extracted from the spectra. This information is complemented with the computation of specific quantities to quantify the amount of asymmetry ($RSk$, relative skewness) of the lines considered for the determination of the line-broadening parameters. 

We identify a variety of line-profiles in terms of shape and broadening characteristics which are quantified in terms of two line-broadening parameters. Following \cite{Sim14}, we divide the sample in two main groups (comprising three subgroups each) based on the ratio \vmac/\vsini\ and taking into account the limitations of the methodology considered for the determination of these parameters. The first group include stars showing an important (or dominant) macroturbulent broadening contribution to the line-profile. The second group consists of stars in which either rotational broadening dominates or both \vsini\ and \vmac\ are below the limits of reliability of the methodology. We also identify cases in which only upper limits of \vmac\ can be determined. These mainly refer to line-profiles dominated by rotational broadening. In this case, the derived values of \vmac\ (which can indeed be large) must be considered with caution.

We then locate the complete sample in the spectroscopic HR diagram separating the stars by line-broadening characteristics. We find that the stars with an important contribution of macroturbulent broadening to their line-profiles are mainly concentrated above $\approx$15~$M_{\odot}$. From the distribution of line-broadening properties in the sHR diagram we find empirical evidence suggesting the existence of various types of non-rotational broadening agents acting in the realm of massive stars. Even though all of them could be quantitatively characterized in terms of \vmac\ -- and quoted as macroturbulent broadening --, their physical origin can be actually different. Indeed, it is natural to think that some of them could be acting at the same time, with a different relative contribution to the line-profiles, depending on the specific properties of the star at a given moment during its evolution. In particular, under this scenario, the distribution of stars in the sHR diagram indicates that, on the one hand, below  $\approx$15~M$_{\odot}$ (or, equivalently log~$\mathscr{L}/\mathscr{L}_{\odot}\approx$3.5) the final shape of the stellar lines depends on a combination of different factors which are not only controlled by \Teff\, \grav, and/or \vsini. On the other hand, the homogeneity in the type of profiles found in stars in the upper part of the sHR diagram ($\gtrsim$15~M$_{\odot}$) calls for one dominant broadening agent which, indeed, is operating in a very similar way in stars with similar $L$/$M$ but very different evolutionary stages.

The investigation of the distribution of single snapshot line-profile properties of our sample of O and B stars in the $RSk$\,--\,\vmac/\vsini\ diagram seems to indicate a potential connection between macroturbulent broadening, line-asymmetries and a variable phenomenon which is producing a time-dependent distribution of motions in the line formation region with an average velocity close to zero. This statement must, however, be confirmed using multi-epoch observations, since we have found that the effect of noise in the measurement of the quantity $RSk$ may also partly explain the observed distribution. 

Following this idea of using multi-epoch observations to provide further observational clues to understand macroturbulent broadening in massive stars, we started a few years ago the compilation of a spectroscopic dataset specifically designed to this aim. While awaiting for the complete dataset, we present in this paper a proof-of-concept investigation of a sample of eight O and B star for which we already count on long time-series observations gathered during several years, including $\approx$~30\,--\,170 spectra depending on the target. This sample comprises four O stars and B~Sgs with still unknown pulsational properties (if any), and four of the well known pulsating/spotted B main sequence stars considered by \cite{Aer14}. A preliminary comparative characterization of the global shape and type of variability found in this small sample of stars seems to also go in the direction of a unique (or dominating) broadening agent acting in O stars and B~Sgs which may not be related to spots or the type of oscillations found in less massive B main sequence stars (commonly associated with pressure and/or gravity heat-driven modes). In addition, we find observational evidence indicating that the line-broadening and line-profile variability observed in O stars and B~Sgs may not be necessarily connected with the same physical driver. 

Based on all the empirical results presented along this paper, and the various scenarios proposed to date to explain the occurrence of macroturbulent broadening and line-profile variability in O and B stars, we launch the following hypotheses: 
\begin{itemize}
\item Surface motions initiated by turbulent pressure instabilities generated in sub-surface convection zones are the main responsible for the dominant non-rotational line-broadening component in O stars and B~Sgs \cite[as already pointed out by][]{Gra15a},
\item In the B main sequence domain, once the effect of vigorous sub-surface convection becomes negligible, stellar oscillations associated with heat-driven modes become important broadening agents \cite[see also][]{Aer14}, especially when the projected rotational velocity of the star is low. However, these sources of line-broadening never lead values of \vmac\ as large as in the case of more massive stars.
\item The most clearly detected line-profile variability is mainly produced by heat-driven pulsational modes, probably in combination with other mechanisms giving rise to spectroscopic variability, such as spots, wind variability, strange modes ... 
\end{itemize}
All together, these ideas provide a scenario which explains most of the observational properties of the line-profiles considered in this paper. However, before reaching a firm conclusion, it must still be further evaluated using 
\begin{itemize}
 \item [(a)] time-dependent spectroscopy of a larger sample of targets,
 \item [(b)] theoretical simulations to confirm that the turbulent pressure scenario may end up in the observed broadening,
 \item [(c)] results from the computation of instability domains of heat-driven high-degree modes for various assumptions on the considered opacities, metal mixture, overshooting parameter, or even on the stellar evolutionary code used as baseline for the adiabatic and non-adiabatic computations, and
 \item [(d)] an extension of the work by \cite{Aer15} to the O star and B~Sg domain
\end{itemize}
In addition, the extension of the type of work presented here to other metallicities (e.g. considering a similar sample of stars in both Magellanic Clouds), and other stellar domains in the HR diagram \cite[see, e.g.][]{Gra15b} will certainly provide interesting clues to further support/dismiss scenarios about macroturbulent broadening in a more general context.

\begin{acknowledgements}
  This work has been funded by the Spanish Ministry of Economy and
  Competitiveness (MINECO) under the grants AYA2010-21697-C05-04,
  AYA2012-39364-C02-01, and Severo Ochoa SEV-2011-0187, and by the Research
  Council of KU\,Leuven under grant GOA/2013/012. Part of this research received
  funding from the European Research Council under the European Community's
  H2020 Framework Programme, grant agreement n$^\circ$670519 (MAMSIE).  This
  paper made use of the IAC Supercomputing facility HTCondor
  (http://research.cs.wisc.edu/htcondor/), recently expanded and improved thanks
  to FEDER funds granted by the Ministry of Economy and Competitiveness, project
  code IACA13-3E-2493. We thank Pieter Degroote for interesting discussions
  related to the work, and the anonymous referee for highlighting the most critical
  points of the first version of this paper.
 
\end{acknowledgements}

%

\begin{thebibliography}{}
\bibitem[Aerts et al.(1992)]{Aer92} Aerts, C., de Pauw, M., \& Waelkens, C.\ 1992, \aap, 266, 294 
\bibitem[Aerts et al.(1998)]{Aer98} Aerts, C., De Cat, P., Cuypers, J., et al.\ 1998, \aap, 329, 137 
\bibitem[Aerts et al.(2009)]{Aer09} Aerts, C., Puls, J., Godart, M., \& Dupret, M.-A.\ 2009, \aap, 508, 409 
\bibitem[Aerts et al.(2010a)]{Aer10a} Aerts, C., Lefever, K., Baglin, A., et al.\ 2010a, \aap, 513, L11 
\bibitem[Aerts et al.(2010b)]{Aer10b} Aerts, C., Christensen-Dalsgaard, J., \& Kurtz, D.~W.\ 2010, Asteroseismology, Astronomy and Astrophysics Library.~ISBN 978-1-4020-5178-4.~Springer Science+Business Media B.V., 2010b
\bibitem[Aerts et al.(2014)]{Aer14} Aerts, C., Sim{\'o}n-D{\'{\i}}az, S., Groot, P.~J., \& Degroote, P.\ 2014, \aap, 569, A118 
\bibitem[Aerts \& Rogers(2015)]{Aer15} Aerts, C., \& Rogers, T.~M.\ 2015, \apjl, 806, L33 
\bibitem[Bailey et al. (2015)]{Bailey2015} Bailey, J.~E., Nagayama, T., Loisel, G.~P., et al.\ 2015, Nature, 517, 56B
\bibitem[Balona(1986)]{Bal86} Balona, L.~A.\ 1986, \mnras, 219, 111 
\bibitem[Bouret et al.(2012)]{Bou12} Bouret, J.-C., Hillier, D.~J., Lanz, T., \& Fullerton, A.~W.\ 2012, \aap, 544, A67 
\bibitem[Briquet et al.(2004)]{Bri04} Briquet, M., Aerts, C., L{\"u}ftinger, T., et al.\ 2004, \aap, 413, 273 
\bibitem[Cantiello et al. (2009)]{Can09} {Cantiello}, M., {Langer}, N., {Brott}, I. et al.\ 2009, \aap, 499, 279
\bibitem[Castro et al.(2012)]{Cas12} Castro, N., Urbaneja, M.~A., Herrero, A., et al.\ 2012, \aap, 542, AA79 
\bibitem[Castro et al.(2014)]{Cas14} Castro, N., Fossati, L., Langer, N., et al.\ 2014, \aap, 570, LL13 
\bibitem[Conti \& Ebbets(1977)]{Con77} Conti, P.~S., \& Ebbets, D.\ 1977, \apj, 213, 438 
\bibitem[Cuypers et al.(2002)]{Cuy02} Cuypers, J., Aerts, C., Buzasi, D., et al.\ 2002, \aap, 392, 599 
\bibitem[De Cat \& Aerts(2002)]{Dec02} De Cat, P., \& Aerts, C.\ 2002, \aap, 393, 965 
\bibitem[Dufton et al.(2006)]{Duf06} Dufton, P.~L., Ryans, R.~S.~I., Sim{\'o}n-D{\'{\i}}az, S., Trundle, C., \& Lennon, D.~J.\ 2006, \aap, 451, 603 
\bibitem[Ekstr{\"o}m et al.(2012)]{Eks12} Ekstr{\"o}m, S., Georgy, C., Eggenberger, P., et al.\ 2012, \aap, 537, A146 
\bibitem[Fraser et al.(2010)]{Fra10} Fraser, M., Dufton, P.~L., Hunter, I., \& Ryans, R.~S.~I.\ 2010, \mnras, 404, 1306 
\bibitem[Fossati et al.(2015)]{Fos15} Fossati, L., Castro, N., Sch{\"o}ller, M., et al.\ 2015, \aap, 582, A45 
\bibitem[Gray(1976)]{Gra76} Gray, D.~F.\ 1976, Research supported by the National Research Council of Canada.~New York, Wiley-Interscience, 1976.~484 p.,  
\bibitem[Gray(1978)]{Gra78} Gray, D.~F.\ 1978, \solphys, 59, 193 
\bibitem[Grassitelli et al.(2015a)]{Gra15a} Grassitelli, L., Fossati, L., Sim{\'o}n-Di{\'a}z, S., et al.\ 2015a, \apjl, 808, L31 
\bibitem[Grassitelli et al.(2015b)]{Gra15b} Grassitelli, L., Fossati, L., Langer, N., et al.\ 2015b, \aap, 584, L2 
\bibitem[Grevesse \& Noels(1993)]{Grevesse1993} {Grevesse}, N. \& {Noels}, A. 1993, in Perfectionnement de l'Association Vaudoise des Chercheurs en Physique, ed. {B.~Hauck, S.~Paltani, \& D.~Raboud}, 205--257
\bibitem[Godart(2011)]{Godart2011} Godart, M. 2011, PhD thesis, University of Liege, Belgium
\bibitem[Godart et al. (2016)]{God16} Godart, M., Sim\'on-D\'iaz, S., Herrero, A. 2016, \aap, submitted
\bibitem[Hekker \& Aerts(2010)]{Hek10} Hekker, S., \& Aerts, C.\ 2010, \aap, 515, A43 
\bibitem[Howarth(2004)]{How04} Howarth, I.~D.\ 2004, Stellar Rotation, 215, 33 
\bibitem[Iglesias \& Rogers(1996)]{Iglesias1996} Iglesias, C.~A. \& Rogers, F.~J. 1996, \apj, 464, 943
\bibitem[Langer \& Kudritzki(2014)]{Lan14} Langer, N., \& Kudritzki, R.~P.\ 2014, \aap, 564, AA52 
\bibitem[Lefever et al.(2007)]{Lef07} Lefever, K., Puls, J., \& Aerts, C.\ 2007, \aap, 463, 1093 
\bibitem[Lefever et al.(2010)]{Lef10} Lefever, K., Puls, J., Morel, T., et al.\ 2010, \aap, 515, A74 
\bibitem[Lucy(1976)]{Luc76} Lucy, L.~B.\ 1976, \apj, 206, 499 
\bibitem[Markova \& Puls(2008)]{Mar08} Markova, N., \& Puls, J.\ 2008, \aap, 478, 823 
\bibitem[Markova et al.(2014)]{Mar14} Markova, N., Puls, J., Sim{\'o}n-D{\'{\i}}az, S., et al.\ 2014, \aap, 562, A37 
\bibitem[Martins et al.(2015)]{Mar15} Martins, F., Herv{\'e}, A., Bouret, J.-C., et al.\ 2015, \aap, 575, A34 
\bibitem[Mahy et al.(2015)]{Mah15} Mahy, L., Rauw, G., De Becker, M., Eenens, P., \& Flores, C.~A.\ 2015, \aap, 577, A23 
\bibitem[Martins \& Palacios(2013)]{Mar13} Martins, F., \& Palacios, A.\ 2013, \aap, 560, A16 
\bibitem[Miglio et al.(2007a)]{Miglio2007a} {Miglio}, A., {Montalb{\'a}n}, J., \& {Dupret}, M.-A. 2007a, \mnras, 375,  L21 -- L25
\bibitem[Miglio et al.(2007b)]{Miglio2007b} {Miglio}, A., {Montalb{\'a}n}, J., \& {Dupret}, M.-A. 2007b, CoAst, 151,  48 -- 56
\bibitem[Moravveji et al.(2012a)]{Mor12a} Moravveji, E., Guinan, E.~F., Shultz, M., Williamson, M.~H., \& Moya, A.\ 2012a, \apj, 747, 108 
\bibitem[Moravveji et al.(2012b)]{Mor12b} Moravveji, E., Moya, A., \& Guinan, E.~F.\ 2012b, \apj, 749, 74 
\bibitem[Moravveji(2016)]{Mor16} Moravveji, E.\ 2016, \mnras, 455, L67 
\bibitem[Noels \& Scuflaire (1986)]{Noe86} Noels, A. \& Scuflaire, R.\ 1986, \aap, 161, 125\bibitem[Pamyatnykh(1999)]{Pam99} Pamyatnykh, A.~A.\ 1999, \actaa, 49, 119
\bibitem[Puls et al.(2005)]{Pul05} Puls, J., Urbaneja, M.~A., Venero, R., et al.\ 2005, \aap, 435, 669 
\bibitem[Raskin et~al.(2011)]{Ras11} Raskin, G., van Winckel, H., Hensberge, H., et~al.\ 2011, \aap, 526, A69 
\bibitem[Reed(2003)]{Ree03} Reed, B.~C.\ 2003, \aj, 125, 2531 
\bibitem[Rivero Gonz{\'a}lez et al.(2012)]{Riv12} Rivero Gonz{\'a}lez, J.~G., Puls, J., Najarro, F., \& Brott, I.\ 2012, \aap, 537, A79 
\bibitem[Rogers et al.(2013)]{Rog13} Rogers, T.~M., Lin, D.~N.~C., McElwaine, J.~N., \& Lau, H.~H.~B.\ 2013, \apj, 772, 21 
\bibitem[Ryans et al.(2002)]{Rya02} Ryans, R.~S.~I., Dufton, P.~L., Rolleston, W.~R.~J., et al.\ 2002, \mnras, 336, 577
\bibitem[Sab{\'{\i}}n-Sanjuli{\'a}n et al.(2014)]{Sab14} Sab{\'{\i}}n-Sanjuli{\'a}n, C., Sim{\'o}n-D{\'{\i}}az, S., Herrero, A., et al.\ 2014, \aap, 564, A39 
\bibitem[Saio et~al. (1998)]{Saio1998} {Saio}, H., {Baker}, N.~H., \& {Gautschy}, A.\ 1998, \mnras, 294, 622
\bibitem[Salmon et al. (2012)]{Salmon2012} Salmon, S., Montalb{\'a}n, J., Morel, T., et al.\ 2012 \mnras.422, 3460S
\bibitem[Samadi et al.(2010)]{Sam10} Samadi, R., Belkacem, K., Goupil, M.J., et al.\ 2010, \apss, 328, 253
\bibitem[Santolaya-Rey et al.(1997)]{San97} Santolaya-Rey, A.~E., Puls, J., \& Herrero, A.\ 1997, \aap, 323, 488 
\bibitem[Shiode et al.(2013)]{Shi13} Shiode, J.~H., Quataert, E., Cantiello, M., \& Bildsten, L.\ 2013, \mnras, 430, 1736 
\bibitem[Schrijvers et al.(1997)]{Sch97} Schrijvers, C., Telting, J.~H., Aerts, C., Ruymaekers, E., \& Henrichs, H.~F.\ 1997, \aaps, 121, 343
\bibitem[Sim{\'o}n-D{\'{\i}}az(2015)]{Sim15a} Sim{\'o}n-D{\'{\i}}az, S.\ 2015, IAU Symposium, 307, 194 
\bibitem[Sim{\'o}n-D{\'{\i}}az \& Herrero(2007)]{Sim07} Sim{\'o}n-D{\'{\i}}az, S., \& Herrero, A.\ 2007, \aap, 468, 1063 
\bibitem[Sim{\'o}n-D{\'{\i}}az \& Herrero(2014)]{Sim14} Sim{\'o}n-D{\'{\i}}az, S., \& Herrero, A.\ 2014, \aap, 562, 135 
\bibitem[Sim{\'o}n-D{\'{\i}}az et al.(2010)]{Sim10} Sim{\'o}n-D{\'{\i}}az, S., Herrero, A., Uytterhoeven, K., et al.\ 2010, \apjl, 720, L174 
\bibitem[Sim{\'o}n-D{\'{\i}}az et al.(2011a)]{Sim11a} Sim{\'o}n-D{\'{\i}}az, S., Castro, N., Garcia, M., et al.\ 2011a, BSRSL, 80, 514 
\bibitem[Sim{\'o}n-D{\'{\i}}az et al.(2011b)]{Sim11b} Sim{\'o}n-D{\'{\i}}az, S., Castro, N., Herrero, A., et al.\ 2011b, JPhCS, 328, 012021 
\bibitem[Sim{\'o}n-D{\'{\i}}az et al.(2012)]{Sim12} Sim{\'o}n-D{\'{\i}}az, S., Castro, N., Herrero, A., et al.\ 2012, Proceedings of a Scientific Meeting in Honor of Anthony F.~J.~Moffat, 465, 19 
\bibitem[Sim{\'o}n-D{\'{\i}}az et al.(2015)]{Sim15b} Sim{\'o}n-D{\'{\i}}az, S., Negueruela, I., Ma{\'{\i}}z Apell{\'a}niz, J., et al.\ 2015, Highlights of Spanish Astrophysics VIII, 576 
\bibitem[Slettebak(1956)]{Sle56} Slettebak, A.\ 1956, \apj, 124, 173 
\bibitem[Sundqvist et al.(2013a)]{Sun13a} Sundqvist, J.~O., Petit, V., Owocki, S.~P., et al.\ 2013a, \mnras, 433, 2497
\bibitem[Sundqvist et al.(2013b)]{Sun13b} Sundqvist, J.~O., Sim{\'o}n-D{\'{\i}}az, S., Puls, J., \& Markova, N.\ 2013b, \aap, 559, L10 
\bibitem[Telting \& Schrijvers(1997)]{Tel97} Telting, J.~H., \& Schrijvers, C.\ 1997, \aap, 317, 723 
\bibitem[Telting et al.(2006)]{Tel06} Telting, J.~H., Schrijvers, C., Ilyin, I.~V., et al.\ 2006, \aap, 452, 945 
\bibitem[Telting et~al.(2014)]{Tel14} Telting, J.~H., \'Avila, G., Buchhave, L., et~al.\ 2014, AN, 335, 41 
\bibitem[Turck-Chi{\`e}ze \& Gilles(2013)]{TurckChieze2013} {Turck-Chi{\`e}ze}, S. \& {Gilles}, D. 2013, EPJ Web of Conferences 43, 01003 
\bibitem[Unno et al.(1989)]{Unn89} Unno, W., Osaki, Y., Ando, H., Saio, H., \& Shibahashi, H.\ 1989, Nonradial oscillations of stars, Tokyo: University of Tokyo Press, 1989, 2nd ed.
\bibitem[Ventura et al.(2008)]{Ventura2008} {Ventura}, P., {D'Antona}, F., \& {Mazzitelli}, I. 2008, \apss, 316, 93
\bibitem[Zdravkov \& Pamyatnykh (2008)]{Zdravkov2008} {Zdravkov}, T. \& {Pamyatnykh}, A.~A. 2008, JPhCS, 118, 012079
\bibitem[Wade et al.(2014)]{Wad14} Wade, G.~A., Grunhut, J., Alecian, E., et al.\ 2014, Magnetic Fields throughout Stellar Evolution, 302, 265 
\end{thebibliography}
%


\begin{appendix} 

\section{Tables}\label{sec-append2}

\end{appendix} 

\end{document}